\documentclass[format=acmsmall, review=false, screen=true]{acmart}

\usepackage{lineno,hyperref}

\usepackage{url}

\usepackage{graphicx}
\usepackage{subfig}

\usepackage{multirow}
\usepackage{booktabs}
\usepackage{tabularx}
\usepackage{ragged2e}
\usepackage{caption}

\usepackage{amsmath}
\usepackage{amsfonts}
\usepackage{euscript}
\usepackage{amssymb}

\usepackage{fixltx2e}

\usepackage{pdflscape}

\usepackage{algorithm}
\usepackage{algpseudocode}

\makeatletter
\def\BState{\State\hskip-\ALG@thistlm}
\makeatother

\algnewcommand\algorithmicforeach{\textbf{for each}}
\algdef{S}[FOR]{ForEach}[1]{\algorithmicforeach\ #1\ \algorithmicdo}

\usepackage{fixltx2e}

\usepackage{pdflscape}

\usepackage{changepage}   

\acmJournal{TAAS}
\acmVolume{0}
\acmNumber{0}
\acmArticle{0}
\acmYear{2018}
\acmMonth{3}

\setcopyright{acmcopyright}

\acmDOI{0000001.0000001}

\received{March 2018}
\received[revised]{March 2018}
\received[accepted]{March 2018}

\begin{document}

\title[A Reference Architecture and Modelling Principles for Architectural Stability]{A Reference Architecture and Modelling Principles for Architectural Stability based on Self-Awareness: Case of Cloud Architectures}

\author{Maria Salama}
\affiliation{%
  \institution{School of Computer Science, University of Birmingham}
  \city{Birmingham}
  \postcode{B15 2TT}
  \country{UK}
}
\email{m.salama@cs.bham.ac.uk}

\author{Rami Bahsoon}
\affiliation{%
  \institution{School of Computer Science, University of Birmingham}
  \city{Birmingham}
  \postcode{B15 2TT}
  \country{UK}
}
\email{r.bahsoon@cs.bham.ac.uk}

\author{Rajkumar Buyya}
\affiliation{%
 \institution{Cloud Computing and Distributed Systems (CLOUDS) Lab, School of Computing and Information Systems, University of Melbourne}
 \city{Melbourne}
 \state{Victoria}
 \postcode{3010 VIC}
 \country{Australia}
}
\email{rbuyya@unimelb.edu.au}

\begin{abstract}
With the increased dependence on software, there is a pressing need for engineering long-lived software. As architectures have a profound effect on the life-span of the software and the provisioned quality of service, stable architectures are significant assets. Architectural stability tends to reflect the success of the system in supporting continuous changes without phasing-out. The \textit{behavioural} aspect of stability is essential for seamless operation, to continuously keep the provision of quality requirements stable and prevent architecture's drifting and phasing-out. In this paper, we introduce a reference architecture and model for stability. Specifically, we leverage on the self-awareness principles and runtime goals modelling to explicitly support architectural stability. To illustrate the applicability and evaluate the proposed approach, we consider the case of cloud architectures. The experimental results show that our approach increases the efficiency of the architecture in keeping the expected behaviour stable during runtime operation.
\end{abstract}

%
%


%
%


\maketitle

\renewcommand\shortauthors{Salama, M. et al}

\section{Introduction}
\label{sec_introduction}
Modern software systems are increasingly operating in highly open, dynamic and uncertain environments \cite{Chen2016}. Such challenges can have impact on the software life-time and the quality of the service provided. This growth, which is likely to continue into the foreseeable future, has motivated the need for long-lived software. An essential prerequisite for longevity of software systems is its capability to maintain service provision with expected qualities and accommodate changes in requirements and environment. 

An extensive literature survey \cite{Salama2018a} has revealed that the stability property has been considered at different levels (e.g. code, design, architecture levels) and with respect to several aspects (e.g. logical, structural, physical). This implies many different interpretations for considering stability as a software property. At the architecture level, stability has been viewed as the ability to endure with changes in requirements and the environment, while reducing the likelihood of architectural drifting and phasing-out, by avoiding ripple structural modifications (over two or more versions the software) \cite{Bahsoon2003b} \cite{Bahsoon2009}. That is an \textit{evolutionary perspective} in considering stability, i.e. evolving the system through a number of releases \cite{Kramer2007}. Meanwhile, dynamic changes, which occur while the system is in operation, require quick and dynamic adaptations during runtime \cite{Kramer2007}. This calls for an \textit{operational perspective} of stability that is fundamental for software architectures, to ensure seamless operation. 

As architectures have a profound effect on the life-span of the software and the quality of service (QoS) provision \cite{Garlan2000} \cite{Garlan2015}, the architecture's behaviour tends to reflect the success of the system in constantly provisioning end-users' requirements, as well as supporting and tolerating continuous changes and evolution over time \cite{Salama2017b}. We argue that architectural stability manifests itself as a software property necessary for the operation of software systems, their dependability and longevity over time \cite{Salama2017b}. To leverage the capabilities of software systems, it is necessary to consider \textit{behavioural stability} to ensure that the architecture's intended behaviour is provisioned during runtime operation. This imposes new questions on how to design and cost-effectively operate such systems. In particular, practitioners and architects are challenged by how they can systematically design for stability, select architecture styles and make design decisions that are stable yet dependable in supporting likely changes in requirements and the environment. 

With the typical key role of architectures in achieving quality requirements \cite{Kazman1994b} \cite{Sommerville2011} \cite{Barber2003} \cite{Ameller2013}, we can evidently agree that realising stability at the architecture level should be based on the quality requirements subject to stability \cite{Kazman1994b} \cite{Ameller2013} \cite{Angelopoulos2013}, where requirements are the key to long-term stability and sustainability \cite{Becker2016} \cite{Chitchyan2016}. In other words, the outward requirements goal is concerned with what the system will accomplish for its end-users \cite{Witt1994}, which will be achieved by the architecture. 

Achieving behavioural stability for long-living software calls for stability planning starting in an early development stage, i.e. in the requirements engineering and architecture design phase \cite{Bass2012}, where stability requirements are assessed throughout the architecture's lifespan and will be used in informing architecture decisions, so that the architecture will not break-down easily when coping with increased runtime load demands or evolution \cite{Lamsweerde2000} \cite{Bahsoon2005}.  Hence, a ``behaviourally stable" architecture design should be based on the requirements subject to stability. Requirements engineering for stability will help in capturing and analysing the quality attributes subject to stability while building stable architectures. Such requirements subject to stability should be modelled as goals at an abstract level, then technically fine-grained to be allocated to single specific components \cite{Letier2002} \cite{Becker2012}. Explicit relation between the requirements model and the architecture should also be present to consider the architectural stability \cite{Nuseibeh2001} \cite{Lamsweerde2000} \cite{Emmerich2002} \cite{Bahsoon2009}. This will result in having the necessary runtime actions to keep the architecture stable, more effective and less costing on the long-term.

Even though architecture design has been widely investigated and derived from quality attributes \cite{Kazman1994a}, stability was not explicitly tackled. The shortcoming of current software engineering practices regarding stability is that the stable provision of certain quality attributes essential for end-users (e.g. response time for real-time systems) is not explicitly considered in requirements modelling and architecture design. To address this problem, we propose a reference architecture and modelling principles for stability based on self-awareness concepts. The main purpose of the work is to facilitate and guide the design of stable architectures for new systems and the improvement of developed systems with architectural stability. 

\textbf{Contributions.} 
The main contributions of our research in this paper are as follows.
\begin{itemize}
\item We employ the quality-driven self-aware and self-adaptive architecture proposed earlier in our previous work \cite{Salama2015} for designing stability-driven architecture pattern\footnote{This paper is an substantially extended version of our conference paper \cite{Salama2015}.}. The pattern leverages on the principles of self-awareness and self-expression \textemdash that have recently emerging in the field of software engineering as a mechanism to seamlessly improve the quality of runtime adaptations, the fulfilment of runtime requirements and the management of complex dynamic trade-offs \cite{Parsons2011}. The proposed architecture incorporates quality self-management generic components and embeds a catalogue of architecture tactics within self-awareness capabilities. Such architecture would take adaptation decisions for better tuning, responding and achieving stability goals. 

\item We present runtime goals modelling for stability featuring novel extensions for the Runtime Goal Models \cite{Dalpiaz2013} \textemdash that is based on the Goal-Oriented Requirements Engineering (GORE) \textemdash in order to enable efficient use of self-awareness and self-expression in achieving stability goals. The extensions include finer-grained and dynamic knowledge representation of the runtime goals, i.e. goals attributes necessary for enabling self-awareness and measures of goals satisfaction in relation to adaptation decisions.

\item We present algorithms for systematic realisation of the symbiotic relation between runtime goals and self-aware architecture pattern, as bidirectional feedback loops interleaved and intertwined, explicitly considering dependencies between stability goals and their corresponding adaptations. The algorithms aim to keep the runtime goals model ``live'' and updated, reflecting on the extent to which the adaptation decisions satisfy stability goals during runtime, promising more accurate and better-informed adaptation decisions, leading to a stable state. The symbiotic realisation is based on dynamic modelling of tactics impact on stability, using Markov analytical model and queueing theory. The premise is that self-awareness can enable the analysis and evaluation for the extent to which candidate tactics can meet stability goals and keep the architecture in stable behaviour.

\item We apply the reference architecture and model to the case of cloud architectures, where the continuous satisfaction and provision of quality requirements without SLA violations in the highly dynamic operating environment are challenging. The cloud architecture is modelled and simulated by extending \textit{CloudSim} \cite{Calheiros2011}. Our work is experimentally evaluated using the \textit{RUBiS} benchmark \cite{RUBiS} varying the number of requests proportionally according to the \textit{World Cup 1998} workload trend \cite{WorldCup98}. Experimental results have shown that the proposed design artefacts have improved the stability in delivering the quality of service goals.
\end{itemize}

The proposed design-support artefact would assist architects and practitioners in planning for stability, as well as designing stable and long-living systems. Such design-support would increase the efficiency of the architecture runtime operation, preventing the architecture from drifting and phasing-out as a consequence of continuous unsuccessful provision of quality requirements. As reference architectures refer to ``a special type of software architecture that have become an important element to systematically reuse architectural knowledge" \cite{Affonso2013}, the reference architecture makes it possible to more systematically design stable architectures. 

\textbf{Organisation.} The rest of the paper is organised as follows. In section \ref{sec_background}, we describe relevant background. In section \ref{sec_stability}, we sketch the properties of architectural stability as a software property. Section \ref{sec_architecture} and \ref{sec_goalModelling} elaborate the technical contributions on reference architecture and goals modelling for stability. Section \ref{sec_relation} presents the symbiotic relation between stability goals and architecture. Section \ref{sec_evaluation} applies our architecture to the case of cloud, followed by experimental evaluation. We discuss the threats to validity of the proposed work and related work in section \ref{sec_threats} and \ref{sec_relatedWork} respectively. Section \ref{sec_conclusion} concludes the paper and indicates future work.

\section{Background}
\label{sec_background}
In this section, we introduce the main concepts (section \ref{sec_background_concepts}). Then, we present an overview on self-awareness (\ref{sec_background_self-awareness}) and Runtime Goal Models (section \ref{sec_background_goalModel}) on which we base our work.

 \subsection{Definitions of the Main Concepts}
\label{sec_background_concepts}

\paragraph{Software Architecture} 
The concept of software architecture has been defined in different ways under different contexts. In our work, we adopt the definition of the ISO/IEC/IEEE Standards that defines software architecture as the ``fundamental organisation of a system embodied in its components, their relationships to each other, and to the environment, and the principles guiding its design and evolution'' \cite{ISO2010Vocab}. This definition is in line with early definitions when the discipline has emerged \cite{Perry1992} \cite{Shaw1996} and with matured ones appearing later \cite{Bass2003}. Software architectures provide abstractions for representing the structure, behaviour and key properties of a software system \cite{Shaw1996}. They are described in terms of software components (computational elements), connectors (interaction elements), their configurations (specific compositions of components and connectors) and their relationship to the environment \cite{Medvidovic2000} \cite{Seo2008}. 

\paragraph{Software life cycle} 
The life cycle of a software system consists basically of the \textit{development} and \textit{operation} phases \cite{Avizienis2004}. The development phase includes all activities till the decision that the software is ready for operation to deliver service, such as requirements elicitation, conceptual design, architectural design, implementation and testing \cite{Avizienis2004}. The operation phase begins when the system is deployed, configured and put into operation to start delivering the actual service in the end-user's environment, cutover issues are resolved, and the product is launched \cite{Avizienis2004} \cite{ISO2010Vocab}. The former phase is known as \textit{initial development} or \textit{design-time}, and the latter is usually referred as \textit{runtime}. After the development and launch of the first functioning version, the software product enters to different cycles of maintenance and evolution stages till reaching the phase-out and close-down \cite{Rajlich2000} \cite{Avizienis2004} \cite{ISO2010Vocab}. During the maintenance stage, minor defects are repaired, while the system functionalities and capabilities are extended in major ways in the evolution stage \cite{Rajlich2000}.

\paragraph{Quality Attribute} 
The definition of a quality attribute we use is of the IEEE Standard for a Software Quality Metrics defining quality attribute as ``a characteristic of software, or a generic term applying to quality factors, quality sub-factors, or metric values'' \cite{IEEE1998Quality}. According to the same standard, a \textit{quality requirement} is defined as ``a requirement that a software attribute be present in software to satisfy a contract, standard, specification, or other formally imposed document'' \cite{IEEE1998Quality}.

\paragraph{Architecturally-significant requirements}
Generally, the architecture should fulfil the software requirements, both functional requirements (what the software has to do) and quality requirements (how well the software should perform) \cite{Witt1994} \cite{Gomaa2010}. Functional requirements are implemented by the individual components, while the quality requirements are highly dependent on the organisation and communication of these components \cite{Sommerville2011}. In the software architecture discipline, the architecturally-significant requirements are considered, as not all requirements have equal effect on the architecture \cite{Lianping2013}. Architecturally-significant requirements are a subset of technically challenging requirements, technically constraining and central to the system's purpose. These requirements have significant influence on the architecture design decisions, as they should be satisfied by the architecture \cite{Lianping2013}. \textit{Architecturally-significant functional requirements} may define the essence of the functional behaviour of the system \cite{Anish2014}, while \textit{architecturally-significant quality requirements} are often technical in nature, such as performance targets \cite{Kazman1994a} \cite{Ameller2013}. This special category of requirements, describing the key behaviours that the system should perform, plays a main role in making architectural decisions and has measurable effect on the software architecture. 

\paragraph{System Behaviour} 
The behaviour of a system is the ``observable activity of the system, measurable in terms of quantifiable effects on the environment whether arising from internal or external stimulus'' \cite{ISO2010Vocab}. This is determined by the state-changing operations the system can perform \cite{ISO2010Vocab}.

\paragraph{Self-adaptive software system} 
In general settings, \textit{to adapt} means ``to change a behaviour to conform to new circumstances'' \cite{Astrom1989}. A self-adaptive software ``evaluates its own behaviour and changes behaviour when the evaluation indicates that it is not accomplishing what the software is intended to do, or when better functionality or performance is possible'' \cite{Laddaga1997} \cite{Oreizy1999} \cite{Cheng2009a}. Intuitively, a self-adaptive system is one that has the capability of modifying its behaviour at runtime in response to changes in the dynamics of the environment (e.g. workload) and disturbances to achieve its goals (e.g. quality requirements) \cite{Meng2001}. Self-adaptive systems are composed of two sub-systems: (i) the managed system (i.e. the system to be controlled), and (ii) the adaptation controller (the managing system) \cite{Villegas2011}. The managed system structure could be either a non-modifiable structure or modifiable structure with/without reflection capabilities (e.g. reconfigurable software components architecture) \cite{Villegas2011}. The controller's structure is a variation of the MAPE-K loop \cite{Villegas2011}.

\subsection{Self-Awareness and Self-Expression}
\label{sec_background_self-awareness}
As self-adaptive software systems are increasingly becoming heterogeneous with dynamic requirements and complex trade-offs \cite{Nya2014}, engineering self-awareness and self-expression is an emerging trend in the design and operation of these systems. Inspired from psychology and cognitive science, the concept of self-awareness has been re-deduced in the context of software engineering to realise autonomic behaviour for software exhibiting these characteristics \cite{Lewis2011} \cite{Faniyi2014}, with the aim of improving the quality of adaptation and seamlessly managing these trade-offs.

The principles of self-awareness are employed to enrich self-adaptive architectures with awareness capabilities. As the architectures of such software exhibit complex trade-offs across multiple dimensions emerging internally and externally from the uncertainty of the operation environment, a self-aware architecture is designed in a fashion where adaptation and execution strategies for these concerns are dynamically analysed and managed at runtime using knowledge from awareness. 

A self-aware computational node is defined as a node that ``possesses information about its internal state and has sufficient knowledge of its environment to determine how it is perceived by other parts of the system'' \cite{Lewis2011} \cite{Faniyi2014}. A node is said to have \textit{self-expression} capability ``if it is able to assert its behaviours upon either itself or other nodes, this behaviour is based upon a nodes sense of its personality'' \cite{Parsons2011} \footnote{Architecting self-aware software has been introduced in \cite{Faniyi2014} and detailed in \cite{Chen2014a}}. Different levels of self-awareness, called capabilities, were identified to better assist the self-adaptive process \cite{Parsons2011} \cite{Faniyi2014}: 
\begin{itemize}
	\item \textit{Stimulus-awareness}: a computing node is stimulus-aware when having knowledge of stimuli, enabling the system's ability to adapt to events. This level is a prerequisite for all other levels of self-awareness. 
	\item \textit{Goal-awareness}: if having knowledge of current goals, objectives, preferences and constraints, in such a way that it can reason about it. 
	\item \textit{Interaction-awareness}: when the node's own actions form part of interactions with other nodes and the environment. 
	\item \textit{Time-awareness}: when having knowledge of historical information and/or future phenomena. 
	\item \textit{Meta-self-awareness}: the most advanced of the self-awareness levels, which is awareness of own self-awareness capabilities.
\end{itemize}

\subsection{Runtime Goal Models}
\label{sec_background_goalModel}
Goal-oriented requirements engineering (GORE) has become a widely used paradigm for elicitation, modelling, analysis and reasoning of systems requirements \cite{Lamsweerde2009} \cite{Heaven2011}. Goals are objectives to be achieved by the system under consideration \cite{Letier2002}, i.e. prescriptive statements of intent whose satisfaction requires the cooperation of different components in software and its environment \cite{Lamsweerde2000} \cite{Becker2012}. Goals range from high-level to fine-grained technical prescriptions that can be assigned as responsibilities to single components \cite{Becker2012} \cite{Letier2002}. Goals, thereby, provide a rationale for requirements and allow tracing low-level details back to high-level concerns \cite{Letier2002}.

Runtime Requirements Models \textemdash denoting requirements models that are used at runtime \textemdash have a key role to support monitoring requirements satisfaction and the consequent adaptations during runtime. Runtime Goal Models, extending design-time goals, were proposed to analyse the runtime behaviour of a system with respect to the satisfaction of requirements and consequently refine the goals specification model, its assumptions and operationalisation decisions \cite{Dalpiaz2013} \cite{Souza2011} \cite{Blair2009}. Runtime goals were employed in self-adaptive software catering for uncertainty \cite{Feather1998}. 

\section{Architectural Stability}
\label{sec_stability}
Generally, the notion of ``stability'' refers to the resistance to change and the tendency to recover from perturbations. The condition of being stable, thus, implies that certain properties of interest do not (very often) change relative to other things that are dynamically changing. As a software quality property, stability is defined in the ISO/IEC 9126 standards for software quality model \cite{ISO2000Quality} as one of the sub-characteristics of the maintainability characteristic of the software \textemdash along with analysability, changeability and testability \textemdash as ``the capability of the software product to avoid unexpected effects from modifications of the software'' \cite{ISO2000Quality}. For general application purposes, the standard does not determine specific features or aspects for stability \cite{Dobrica2002}. 

Reviewing the state-of-the-art in software engineering \cite{Salama2018a}, we have found that stability has been considered at different levels, i.e. at the code level (e.g. \cite{Yau1980}), requirements (e.g. \cite{Bush2003}), design (\cite{Yau1985} \cite{Fayad2001} \cite{Elish2005} \cite{Kelly2006}) and the architecture level (\cite{Jazayeri2002} \cite{Bahsoon2003b} \cite{Tonu2006} \cite{Molesini2010}). At each level, stability has been considered in relation to several aspects from different perspectives, and thus interpreted in many ways according to the perspective of consideration. For instance, stability at the code level has been interpreted as ``the resistance to the potential ripple effect that the program would have when it is modified'' \cite{Yau1980}, that is considering the \textit{logical} and performance (i.e. \textit{behavioural}) aspects of stability from the \textit{maintenance} perspective. Design stability has been refered to ``the extent to which the structure of the design is preserved throughout the evolution of the software from one release to the next'' \cite{Yau1985}, where the \textit{logical} and \textit{structural} aspects of stability are considered from \textit{evolutionary} perspective.

Architectural stability has been considered in terms of ripple structural modifications over two or more versions of the software, as a \textit{structural} aspect with respect to architecturally-relevant changes carried from \textit{evolutionary} (\cite{Jazayeri2002} \cite{Bahsoon2003b}) and maintenance perspectives (\cite{Molesini2010}). This has been referred to the extent to which the architecture's structure is capable to accommodate the evolutionary changes without re-designing the architecture or making ripple modifications \cite{Jazayeri2002} \cite{Bahsoon2003b}. Among different perspectives, the \textit{structural} aspect of stability is the one mostly considered at the architecture level.

Considering runtime dynamics of software systems, the structure and the behaviour of the software may be affected when adaptations are taking place during runtime \cite{Chomchumpol2015}. In this context, we distinguish between the \textit{structural} and \textit{behavioural} aspects of stability. We also posit that an \textit{operational} perspective (for the runtime operation of the software) is essential, different from the \textit{evolutionary} perspective (over two or more versions of the software). The stability meaning, we are seeking, can be regarded at the \textit{architecture} level considering the \textit{behavioural} aspect from an \textit{operational} perspective. As such, we define stability as \textit{the ability of the architecture's behaviour to maintain a fixed level of operation (or recover from operational perturbations) within specified tolerances under varying external conditions}. A stable architecture from the \textit{operational} perspective is the one capable to continuously fulfil the architecturally-significant quality requirements during runtime, where the architecture can return to the equilibrium state, following a perturbation due to changes in quality requirements, workload patterns or in the operational environment. Conversely, an unstable architecture is one that, when perturbed from equilibrium, will show deviation from the expected behaviour. So, stability of the architecture is essential to examine the behaviour with time following a perturbation during runtime. 
 
Achieving architectural stability during runtime should start earlier during the design phase. A good plan architecting for stability during design-time will result in keeping the architecture behaviourally stable during runtime, more effective and less costing. Contrarily, putting the architecture in operation with less planning for stability during design-time might lead to instability states that will require unnecessary extra overhead during runtime for keeping the architecture stable. Such instabilities, on the long-term, will end up with drifting and phasing-out the architecture, as the architecture will not be fulfilling the intended behaviour.

\section{Self-aware Reference Architecture for Stability}
\label{sec_architecture}
As the architecture design plays an essential role in delivering the quality requirements \cite{Bass2012}, architectural behavioural stability is directly related to the intended behaviour of the architecture. As an example of behaviour, one architecture could be intended to keep the response time stable (as it is a crucial quality attribute for the end-users in the case of real-time systems), while throughput could be a critical requirement attribute to be kept stable for another architecture. Having the architecture's intended behaviour stable, by assuring the delivery of some quality attributes, is highly desirable. 

The reference architecture is based on an architecture pattern enriched with self-awareness and self-expression components, quality self-management components and catalogue of architectural adaptation tactics for achieving the intended behaviour. Self-awareness capabilities are employed to safeguard the stability of these attributes, where the selection of the appropriate tactic leading to stability will be performed during runtime by the awareness capabilities. Incorporating the tactics, as adaptation actions to meet the quality requirements, will improve and enrich the quality of self-expression, i.e. the adaptation actions taken during runtime \cite{Salama2015}. Such reference architecture allows instantiation of different patterns suitable for different software domain applications interested in stability.

Achieving such stable behaviour requires adaptation actions to cope with the runtime changes. Adaptability is known to be the current routine to consider various ``ilities'' \textendash subject to stability \textendash when architecting systems \cite{Garlan2013}. Architecting for adaptability is meant to make adaptability part of the architecture design reviews, by creating a catalogue of adaptability-enhancing design tactics \cite{Garlan2013}. As such, our reference architecture is enriched with a catalogue of architectural tactics as adaptation actions designated to fulfil quality attributes subject to stability. The architectural pattern is also enriched with quality self-management capabilities, in order to achieve the desired behavioural stability \cite{Salama2015}. 

We envision that self-awareness and self-expression are the most convenient capabilities for realising behavioural stability. The self-awareness capabilities, embedded in the architecture pattern, own the necessary knowledge for achieving stability and keeping the stable state. For instance, the stimulus- and goal-awareness could provide knowledge about stability goals relevant to the system. The time-awareness could help with the historical information and/or future phenomena about achieving stability. 

\begin{figure}[!h]
\centering
\includegraphics[width=0.75\textwidth]{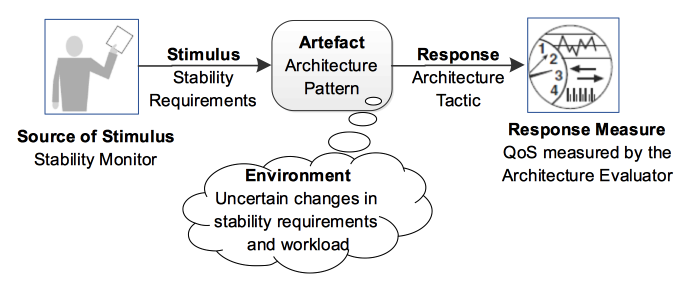}
\caption{General scenario for designing stability-driven pattern (adopted from \cite{Bass2012} \cite{Salama2015})}
\label{fig_scenario}
\end{figure}

To design the reference pattern for achieving stability attributes, we follow the general quality scenario presented at \cite{Bass2012} to formally capture stability requirements. The general scenario, illustrated in Figure \ref{fig_scenario}, is described as follows:

\begin{enumerate}
	\item The Stability Monitor (\textit{source of stimulus}) monitors changes in stability attributes (\textit{stimulus}) during runtime and collect relevant data.
	
	\item The architecture pattern (\textit{artefact}) is responsible for realising stability. Stimulus-awareness is responsible for detecting violations (or possible violations as per threshold) in stability attributes and notifying the self-awareness component to consider adaptation action. The self-awareness responds by selecting an architectural tactic from the Tactics Catalogue, embedded in the pattern, to meet stability requirements and accordingly perform the adaptation actions. 
	
	\item The Self-expression component is, by its turn, responsible for composing the tactic (\textit{response}) and instantiating it as an adaptation action.
	
	\item The \textit{response}, after the execution of the tactic, is measured by the Architecture Evaluator which in turn feeds the different levels of awareness to take further actions if needed and keep history.
\end{enumerate}

\subsection{Quality/Tactics Self-management Generic Components}
\label{sec_architecture_components}
\begin{figure}[!h]
\centering
\includegraphics[width=0.90\textwidth]{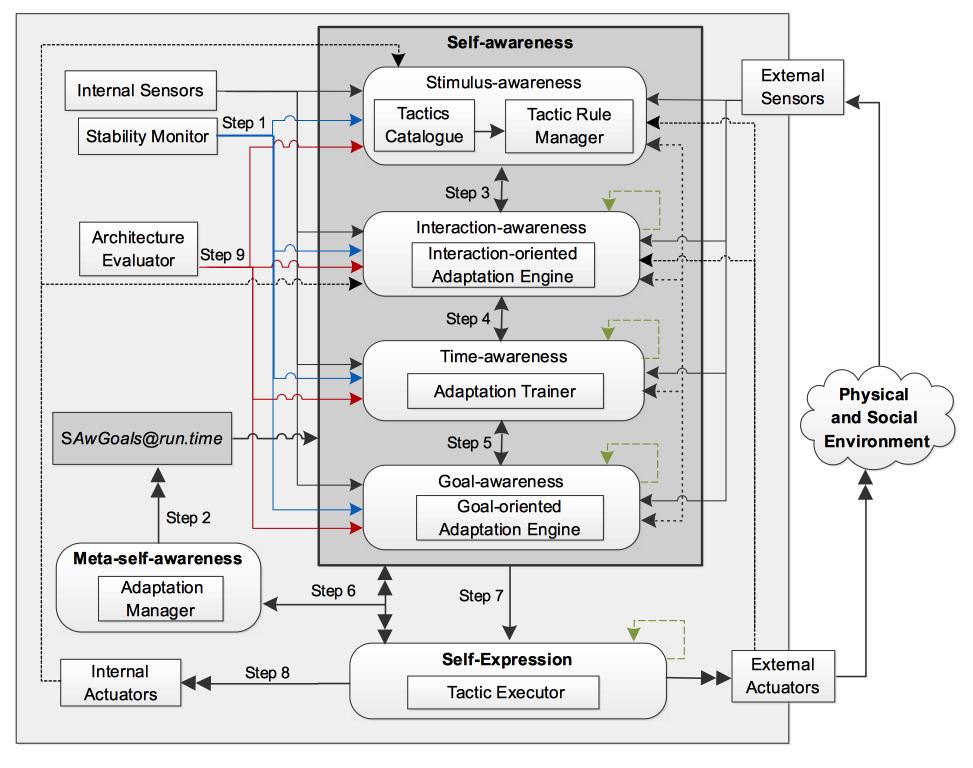}
\caption{Reference architecture pattern with tactics generic components}
\label{fig_metaPattern}
\end{figure}

The reference architecture aims at supporting the process of architecture design for stability. Figure \ref{fig_metaPattern} illustrates the architecture pattern with self-awareness capabilities and tactics generic components. To achieve the envisioned quality self-management capability, the generic components added within self-awareness capabilities are:

\begin{itemize}
	\item \textit{Stability Monitor component}: responsible for monitoring changes in workload and stability attributes during runtime. 
	
	\item \textit{Tactics Catalogue}: a catalogue of runtime tactics designated to achieve different quality attributes subject to stability. As stimulus-awareness is the base of all self-awareness capabilities, the catalogue of architectural tactics is embedded at the stimulus-awareness component.

	\item \textit{Tactics Rule Manager}: embedded at the stimulus-awareness level, it defines \textit{if-condition-then-action} adaptation rules, where the conditions are stability requirements and the actions are response tactics. Rules include priorities for tactics to reflect the order of executing tactics (e.g. vertical scaling is used first before horizontal scaling for faster response and less cost).

	\item \textit{Adaptation Engine}: could be seen as a more complex version of the Tactic Rule Manager present in different levels of awareness. A goal-oriented adaptation engine uses knowledge about design-time and runtime goals available at the goal-awareness component to make decisions about tactic selection in line with the system's current goals. Interaction-oriented adaptation engine contributes to the selection of the adaptation decision according to runtime conditions of the other nodes in the interacting environment where the node is collaborating.

	\item \textit{Adaptation Trainer}: helps in improving the selection of the adaptation decision using historical information. Historical data, received from the Stability Monitor and the Architecture Evaluator, include tactics responses under different runtime conditions to improve the quality and accuracy of adaptation in the future.

	\item \textit{Adaptation Manager}: in the meta-self-awareness level, is responsible for managing trade-offs between stability attributes during runtime and switching between different behavioural strategies in the interaction-, time- and goal-awareness capabilities. The dynamic selection of the appropriate tactic at runtime is performed based on the reasoning about the benefits and costs of selecting a tactic based on a certain level of awareness in order to meet stability attributes while managing trade-offs between them. 

	\item \textit{Tactic Executor}: responsible for managing the process of tactic composition and execution during runtime at the self-expression level. In more details, it makes instructions about the composition and instantiation of the components required to execute the tactic, and the actual execution of the tactic components and connectors during runtime.

		\item \textit{Architecture Evaluator}: evaluates the response after executing of the tactic, and feeds the different levels of awareness to take further actions if needed and accumulate historical information.
\end{itemize}

\subsection{Designing Stability-driven Architecture Patterns}
\label{sec_architecture_patterns}
We discuss how the reference architecture could be instantiated. A variety of patterns could be designed using different combinations of self-awareness capabilities, so that the pattern used when designing the software would include capabilities relevant to the software requirements \cite{Salama2015}. This could follow the methodology for designing self-aware and self-expressive systems proposed in \cite{Chen2014a}. We use the set of self-aware and self-expressive patterns of \cite{Chen2014a} and \cite{Salama2015} that are Basic Pattern (P\textsubscript{1}), Basic Information Sharing Pattern (P\textsubscript{2}), Coordinated Decision-making Pattern (P\textsubscript{3}), Temporal Knowledge Aware Pattern (P\textsubscript{4}), Temporal Knowledge Sharing Pattern (P\textsubscript{5}), Goal Sharing Pattern (P\textsubscript{6}), Temporal Goal Aware Pattern (P\textsubscript{7}), Temporal Goal Sharing Pattern (P\textsubscript{8}), Meta-self-aware Pattern (P\textsubscript{9}), as examples of different possible combinations. The generic components added in different self-aware patterns are summarised in Table \ref{tbl_patterns}.

\begin{table}[!h]
\caption{Variety of stability-driven patterns and their generic components}
\label{tbl_patterns}
\center
\footnotesize
\begin {tabular}{ llllllllll }
\toprule
	\multirow{ 2}{*}{\textbf{Component}}	& 
	\multicolumn{9}{c}{\textbf{Patterns}}
	\\
	\cmidrule(r){2-10}
														&
	\textbf{P\textsubscript{1}}	& 
	\textbf{P\textsubscript{2}}	& 
	\textbf{P\textsubscript{3}}	&
	\textbf{P\textsubscript{4}}	& 
	\textbf{P\textsubscript{5}}	& 
	\textbf{P\textsubscript{6}}	& 
	\textbf{P\textsubscript{7}}	& 
	\textbf{P\textsubscript{8}}	& 
	\textbf{P\textsubscript{9}}	 
	\\
\midrule
	Stability Monitor &	
	$\surd$	&	$\surd$	&	$\surd$	&	$\surd$	&	
	$\surd$	&	$\surd$	&	$\surd$	&	$\surd$	&	$\surd$	
	\\
	Tactics Catalogue	&	
	$\surd$	&	$\surd$	&	$\surd$	&	$\surd$	&	
	$\surd$	&	$\surd$	&	$\surd$	&	$\surd$	&	$\surd$	
	\\
	Tactics Rule Manager	&	
	$\surd$	&	$\surd$	&	$\surd$	&	$\surd$	&	
	$\surd$	&	$\surd$	&	$\surd$	&	$\surd$	&	$\surd$	
	\\
	Goal Adaptation Engine	&	
	-	&	-	&	-	&	-	&	
	- &	$\surd$	&	$\surd$	&	$\surd$	&	$\surd$	
	\\
	Interaction Adaptation Engine		&	
	-	&	$\surd$	&	$\surd$	&	-	&	
	$\surd$	&	$\surd$	&	-	&	$\surd$	&	$\surd$	
	\\
	Adaptation Trainer		&	
	-	&	-	&	-	&	$\surd$	&	
	$\surd$	&	-	&	$\surd$	&	$\surd$ &	$\surd$	
	\\
	Adaptation Manager		&	
	-	&	-	&	-	&	-	&	
	- &	-	&	-	&	-	&	$\surd$	
	\\
	Tactic Executor		&	
	$\surd$	&	$\surd$	&	$\surd$	&	$\surd$	&	
	$\surd$	&	$\surd$	&	$\surd$	&	$\surd$	&	$\surd$	
	\\
	Architecture Evaluator		&	
	$\surd$	&	$\surd$	&	$\surd$	&	$\surd$	&	
	$\surd$	&	$\surd$	&	$\surd$	&	$\surd$	&	$\surd$	
	\\
	\bottomrule
\end{tabular}
\end{table}

\section{Runtime Goals Modelling for Stability}
\label{sec_goalModelling}
In this section, we present the finer-grained knowledge representation of our proposed \textit{SAwGoals@run.time} for modelling runtime stability goals. 

\subsection{Runtime Goals and Self-Awareness}
\label{subsection_goalsAndAwareness}
We propose enriching the architecture pattern with \textit{SAwGoals@run.time} component (as illustrated in Figure \ref{fig_metaPattern}). As runtime goals drive the architecture in reasoning about adaptation during runtime \citep{Sawyer2010}, \textit{SAwGoals@run.time} extends the GORE model to suit the needs of self-awareness capabilities and stability requirements. The objectives of the proposed modelling are: (i) fine-grained dynamic knowledge representation of stability goals to enable efficient use of the different levels of self-awareness, (ii) monitoring the satisfaction of stability goals and the performance of tactics, (iii) better informed decision of the optimal tactic for realising architectural stability, and (iv) continuous accumulation of historical information to update the knowledge for future learning using time-awareness.

We refine the Runtime Goal Models with fine-grained dynamic knowledge representation that reflects self-awareness needs for new attributes of the goals, operationalisation, tracing down to architecture and runtime satisfaction measures. Specifically, additional runtime behavioural details relevant to different levels of self-awareness are integrated, such as node information for interaction-awareness, and trace history for time-awareness, as well as information about the execution environment in different time instances. Operationalisation of stability attributes is realised by self-expression, through runtime tactics which are defined within the proposed model. The model would better operate in the presence of historical information about the ability of operationalisation decisions. In the case of instantiation, it is imperative that the designer consider what-if analysis, simulation or scenarios to test the suitability of the choice. Models which rely on decision-making under uncertainty can also be sensible to employ. Given relevant information about goals and the operating environment, conflict management between goals during runtime is handled by meta-self-awareness capabilities.

The proposed \textit{SAwGoals@run.time} overcomes the limitations of GORE with respect to self-awareness and self-expression as follows:

\begin{itemize}
\item \textit{Goal Attributes}. 
Operating different levels of self-awareness requires detailed information about the goals during runtime. Such information should include attributes about the interacting node, time instance, the execution traces, the adaptations and their performance to satisfy the goal, as well as the operating environment. For instance, information about goals from other nodes and adaptations taking place in the operating environment are required for the interaction-awareness level. Having this information for different time instances would form historical information useful for the time-awareness level to improve the accuracy of adaptation. 

\item \textit{Goal Operationalisation}. 
Operationalisation is performed at the self-expression level using Runtime Goal Model operationalisation, as follows. For operationalising stability attributes, we extend the Runtime Goal Model to introduce alternative of runtime tactics, designed to stabilise and operationalise changes in stability goals at runtime. QoS provision under runtime uncertainty could be handled using alternative operationalisation strategies/ tactics designated for various quality attributes \cite{Bass2003} \cite{Mirakhorli2012}. For instance, self-aware systems encounter during runtime uncertain changes in stability goals due to the changing workloads and size of jobs from users with different SLAs. Runtime tactics designed for performance, like vertical and horizontal scaling, are candidate artefacts for handling stability goals, from which self-awareness can select the optimal handling tactic. The extent to which goals are satisfied is subject to the choice of the tactic. 
 
\item \textit{Conflict Management}. 
As the system encounter operationalisation decisions during runtime for multiple goals, conflicts are likely to exist. Conflict management in dynamic environments exhibits numerous uncertainties and trade-offs requiring intelligent strategies for negotiating conflicts, prioritising and reconciling decisions. Conflict management, through active negotiation, can rely on information related to historical performance of the tactics in meeting the goals. Negotiation is continuously live in self-aware system, as such: once reconciliation is reached and decision is taken, a \textit{trace} of the decision is monitored for its ability to satisfy the goal and possible dependencies. This information can feed into subsequent cycles of negotiation, with the objective of better resolving conflicts the system learns through self-awareness. 
\end{itemize}

\subsection{Runtime Goals Knowledge Representation}
\label{subsection_knowledge}
Runtime goals in \textit{SAwGoals@run.time} are defined along with an execution trace and traced to runtime tactics for operationalisation. A \textbf{Runtime Goal} (e.g. performance) $ G \in \mathcal{G} $, where $ \mathcal{G} $ is the set of goals in a self-aware and self-expressive node. A goal is defined by the following attributes:

\begin{itemize}
\item \textit{Unique identifier} $ id $ of the goal $ G $.

\item \textit{Definition}. formally and informally defining the goal and its satisfaction in an absolute sense. 

\item \textit{Node identifier} $ N $, the unique identifier of the self-aware node responsible for realising the goal.

\item \textit{Weight} $ w $ to consider the priority of the goal.

\item \textit{Metric} $ M $ a measurable unit (e.g. response time measured in milliseconds) that can be used to measure the satisfaction of the goal while the system is running. 

\item \textit{Objective Functions} $ f(G) $ defines the measures for assessing levels of goal satisfaction with respect to values defined in SLAs of different end-users (e.g. objective functions for performance are response time 15 ms and 25 ms for dedicated and shared clients).

\item \textit{Set of tactics} $ T(G) \in \mathcal{T} $ to be used in case of violation of the goal. The goal semantic is the set of system behaviours, i.e. runtime tactics, that satisfy the goal's formal definition. 
\end{itemize}

A \textbf{Runtime Tactic} $ T \in \mathcal{T} $ (e.g. vertical scaling) is defined as follows:
\begin{itemize}
\item \textit{Unique identifier} $ id $ of the tactic $ T $.

\item \textit{Definition} includes description and informal definition for when to apply the tactic and how to execute it.

\item \textit{Object} in the architecture in which the tactic is executed (e.g. VMs).

\item \textit{Pre-condition} defines the current condition of the operating environment in which the tactic could be applied.

\item \textit{Limits} defines the minimum and maximum limits of the architecture for executing the tactic (e.g. the maximum number of servers).

\item \textit{Functionality} defines how the tactic should be executed.

\item \textit{Post-condition}. This characterises the state of the operating environment after applying the tactic. 

\item \textit{Variantions of the tactics} includes different forms or possible configurations for applying the tactic (e.g. earliest deadline first scheduling, least slack time scheduling).
\end{itemize}

A \textbf{Runtime Goal Instance} $ G(n, t_i) $ is an instance of the runtime goal $ G $ in the self-aware node $ n $ at a certain time instance $ t_i $, and is defined as follows:
\begin{itemize}
\item \textit{Client} $ c $ issuing the service request $ r $.

\item \textit{Objective function} denotes the quality value defined in the SLA of the client $ c $.

\item \textit{Tactic} $ T $ and its configuration executed as an adaptation action to satisfy the goal.

\item \textit{Actual value} $ v $ denotes the degree of satisfaction achieved after the execution of the tactic $ T $ that is measured by the Architecture Evaluator.

\item \textit{Set of environment runtime goals} $ G_e $, that are the goals from other self-aware nodes $ n_x $ running at the same time instance $ t_i $ with which the node $ n $ is interacting, where $ G_e = \{ G_1(n_1, t_i), G_2( n_2, t_i), ..., G_x(n_x, t_i) \} $.

\item \textit{Set of environment runtime tactics} $ T_e $, that are the tactics taking place at the same time instance $ t_i $ in the environment, where $ T_e = \{ T_1, T_2, ..., T_x \} \mbox{ for } \forall \mbox{ } G \in G_e $.
\end{itemize}

For each goal $ G $, \textit{change tuples} are created at different time instances $ t_i $ to form the history of this goal $ \mathsf{H}(G) $ for keeping record of the goal satisfaction and related tactics performance over time. This history shall be used by time-awareness to reason about adaptation actions in the future.

\section{Symbiotic Relation Between Runtime Goals and Self-awareness}
\label{sec_relation}
\begin{figure}[!h]
\centering
\includegraphics[width=0.75\textwidth]{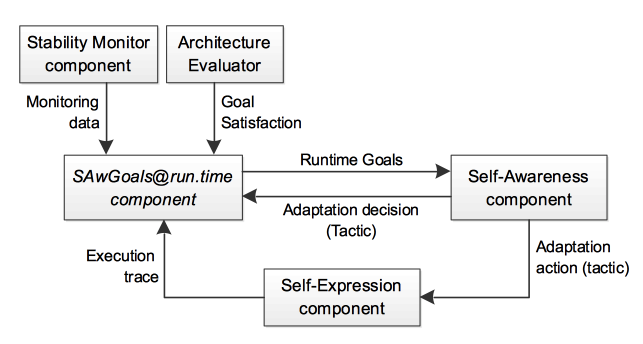}
\caption{Symbiotic relation between Runtime Goals and Self-awareness}
\label{fig_symbioticRelation}
\end{figure}

As end-users requirements change during runtime, there is a need to maintain the synchronisation between the goals model and the architecture \cite{Sawyer2010}. We envision enriching the proposed architecture patterns and goals modelling by incorporating the symbiotic relation between runtime goals and self-awareness capabilities. The symbiotic relation promises more optimal adaptations and better-informed trade-offs management decisions. It aims to keep the runtime goal model ``live'' and up-to-date, reflecting on the extent to which adaptation decisions satisfied the goal(s). The symbiotic relation, illustrated in Figure \ref{fig_symbioticRelation}, is realised during runtime as follows.

\begin{enumerate}
\item Goals are defined and modelled by the \textit{SAwGoals@run.time} component, with fine-grained knowledge representation relevant to the different levels of awareness (as detailed in section \ref{subsection_knowledge}). 

\item Having goals information fed to the self-awareness component, a better informed adaptation decision would be taken based on the learning of time-awareness and the runtime environment of interaction-awareness capabilities.

\item The selected tactic is executed by the self-expression component. 

\item The execution trace is, then, fed back to the goals model to be kept in the log of the goal history. 

\item The goal satisfaction is evaluated by the Architecture Evaluator component to be logged in the goal history. 

\item The goal history is used, in turn, by time-awareness at the next time instance when selecting the appropriate tactic.
\end{enumerate}

\subsection{Algorithms for Realisation}
\label{sec_relation_algorithms}
To realise the symbiotic relation, we provide algorithms to process the Runtime Goal Instance (Algorithm \ref{alg_runtimeGoalInstance}) and construct the Goal History (Algorithm \ref{alg_goalHistory}). 

\textbf{Algorithm 1: Processing Runtime Goal.} This algorithm is launched to process the Runtime Goal Instance $ G(n, t_i) $ at time instance $ t_i $. 

\begin{center}
\begin{minipage}{0.70\linewidth}
\footnotesize
\begin{algorithm}[H]
\caption{Process Runtime Goal}
\label{alg_runtimeGoalInstance}
\begin{algorithmic}[1]
\Procedure {ProcessGoal}{$ G_i = (G_{id}, N_{id}, t_i) $} 
	\State get ObjectiveFunction(client $ c $) 
\BState \emph{QoSMonitor}:
	\State get MonitoringData($ G $)
\BState \emph{Self-awarenessComp}:
	\If {violation($ G $)} 
		\State Identify set of possible tactics $ T(G) $
		\If {TimeAwareness is enabled}
			\State get goal hisotry $ \mathsf{H}(G) $
		\EndIf
		\State select tactic $ T_x \in T(G) $
\BState \emph{Self-expressionComp }:
		\State execute tactic $ T_x $ 
		\State get ExecutionTrace $ \tau(G_i) $ 
\BState \emph{ArchitectureEvaluator}:
		\State get GoalSatisfaction $ v(G) $ 
	\EndIf
\EndProcedure
\end{algorithmic}
\end{algorithm}
\end{minipage}
\par
\end{center}

\textbf{Algorithm 2: Constructing Goal History.} 
This algorithm constructs a change tuple for the goal $ G $ at each time instance $ t_i $. Each change tuple records a log of the objective function, goals from the environment, set of tactics executed in the environment, the tactic executed, the execution trace and the goal satisfaction measure. These change tuples would form the goal history over the different time instances.

\begin{center}
\begin{minipage}{0.75\linewidth}
\footnotesize
\begin{algorithm}[H]
\caption{Construct Goal History}
\label{alg_goalHistory}
\footnotesize
\begin{algorithmic}[1]
\Procedure {ConstructHistory}{Goal $ G = (G_{id}, N_{id}) $}
	\ForEach {$ t_i $}
		\State log time instance $ t_i $
		\State log ObjectiveFunction(client $ c $)
		\State log executed Tactic $ T_x $
		\State log ExecutionTrace $ \tau(G) $
		\State get GoalSatisfaction $ v(G) $ 
		\If {InteractionAwareness is enabled}
			\State log Environment Goals $ G_e = \{ G_1(n_1, i), G_2( n_2, i), ..., G_x(n_x, i) \} $
			\State log Environment Tactics $ T_e = \{ T_1, T_2, ..., T_x \} \mbox{ for } \forall \mbox{ } G \in G_e $
		\EndIf	
	\EndFor
\EndProcedure
\end{algorithmic}
\end{algorithm}
\end{minipage}
\par
\end{center}

\subsection{Dynamic Modelling of Tactics Impact on Stability}
\label{sec_relation_modelling}
A dynamic system exhibits probabilistic behaviour during runtime. Such behaviour is mainly due to the uncertain fluctuation of workload at runtime, the constraints on available resources and changes in the environment. Behaviour can also be affected by prior decisions and adaptation actions. Given the runtime dynamics and the probabilistic behaviours of such system, a Markovian analytical modelling can provide a generic and scalable model for this probabilistic behaviour. Based on multiple parallel dynamic queues, the model can capture instance-related information at finer-grained level of tactics' configurations, given the environment heterogeneity. The model can, then, measure and predict quality attributes for a scenario of interest. Such measurements and predictions, in conjunction with the goal-awareness capability, can assist in choosing the optimal tactics and their configurations to achieve behavioural stability.

\subsubsection{System Model}
\begin{figure}[!h]
\centering
\includegraphics[width=0.75\textwidth]{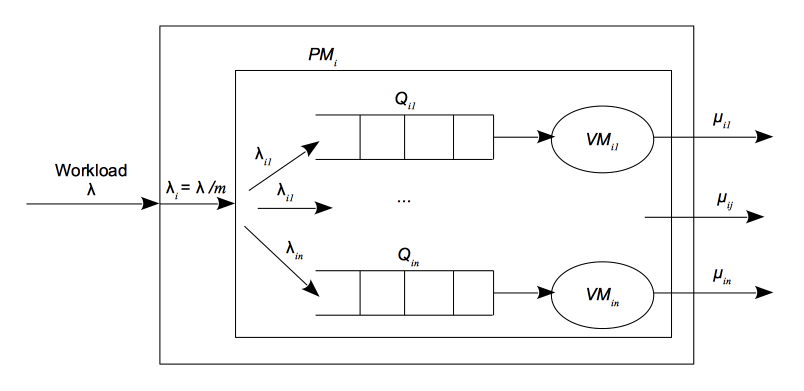}
\caption{Dynamic self-aware workload handling}
\label{fig_systemModel}
\end{figure}

The handling of workload in a self-aware node is illustrated in Figure \ref{fig_systemModel}. Assume that a software system is running on a computing node using $m$ hosts (Physical Machines PMs). A PM$_i$, where $ i = \left \{ 1,..., m \right \} $ runs $n_i$ VMs sharing computational resources. The number of running VMs varies from one PM to another according to its computational capacity. Service requests are received and processed on the infrastructure, where the workload tends to vary in number of incoming requests, length of each request, and quality requirements according to the client SLA. 

We assume the total incoming workload $ \lambda $ will be divided among the $m$  PMs resulting $ \left \{ \lambda_1, \lambda_2, \lambda_i, ..., \lambda_m \right \}  $. Several algorithms have been proposed to manage the jobs placement in PMs and VMs \cite{Yean-Fu2014} \cite{Paul2011}. Though we follow a simple approach for requests placement, the same principle can apply to other placement mechanisms. The distribution of workload, in our case, is based on either the PM computational capacity in case PMs computational capacity are different, or equally on all PMs based on their availability. 

Each PM, but its turn, will distribute its workload share on its $n$ running VMs. Workload is distributed on VMs level either based on VM computational capacity in case the incoming request is constrained by certain computational requirements, or equally in case of no constraints. The workload is denoted by $ \lambda_{ij} $, where $i$ indicates the PM, $j$ indicates the VM, and $ j = \left \{ 1,..., n_i \right \} $. For a VM$_{ij}$, an $ m/m/1 $ queue will be formed for the incoming requests to be processed, where the incoming rate of requests constitutes a poisson process of rate $ \lambda_{i/n} $ (assuming equal workload distributed on all VMs), and the service process is markovian exponentially distributed, with parameter $ \mu_{ij} $ and mean $ 1/\mu_{ij} $ that is handled by that VM. Thus, the total service handled by the self-aware node is $ \sum\limits_{i=1}^m \sum\limits_{j=1}^n \mu_{ij} $. 

Unlike most of the prior models that have employed only single queues, we employ multiple parallel dynamic queues, where the queuing can discipline the way we analyse the workload in relation to heterogeneous environments with varying configurations of PMs, VMs, and their computational capacities. The model also features scalability into the analysis, as well as helps in tracking and predicting the behaviour at a given time instance.

For  VM$_{ij}$, the formed queue of incoming requests can be described as a continuous time Markov chain with transition rate matrix 
\[
Q_{ij} = 
\begin{pmatrix}
  -\lambda_{i/n} & \lambda_{i/n} &  & & \\
  \mu_{ij} & -(\mu_{ij}+\lambda_{i/n}) & \lambda_{i/n} & & \\
  & \mu_{ij} &  -(\mu_{ij}+\lambda_{i/n}) & \lambda_{i/n} & \\
  & & & & \ddots \\
\end{pmatrix}
\]
on the state space $ S_{ij} \left \{ 0,1,2,3, ... \right \} $, and the rate from state $ k $ to the state $ k+1 $ is denoted by $ q_{k, k+1} $. Thus, 
$ q_0 = \lambda_{i/n} $ , $ q_{00} = -\lambda_{i/n} $ , and $ q_{01} = \lambda_{i/n} $ . In general, we must have
\[
q_{k,k+1} \geq 0 \mbox{ for all } k \neq k+1 \in S_{ij} 
\]
where $ q_{k,k+1} $ denotes the $ k,k+1^{th} $ diagonal element in the $ Q_{ij} $ matrix.

Let $ X_t $ denote the number of requests in the VM$_{ij}$ queue at time $ t $. If $ X_t = 0 $, then the next event has to be the arrival of a new request, and the time of its arrival is exponential $ \lambda_{i/n} $. At run-time, the next event could be either the arrival of a new request or the departure of the request currently being processed. Thus, the time to the next event is exponentially distributed with the parameter $ \lambda_{i/n} + \mu_{ij} $. Hence, $ q_k = \lambda_{i/n} + \mu_{ij} $, $ q_{k,k+1} = \lambda_{i/n} $, and $  q_{k,k-1} = \mu_{ij}$. So, the probability of the arrival of a new request is $ \lambda_{i/n} /  \left( \lambda_{i/n} + \mu_{ij} \right) $, and the complementary probability $ \mu_{ij} /  \left( \lambda_{i/n} + \mu_{ij} \right) $ is the probability of the departure of the request currently being processed. 

Having fully specified the transition rate matrix $ Q_{ij} $, $ \left \{ X_t, t \geq 0 \right \} $ is, then, a Markov process with the following transition rates:
\[
\begin{split}
q_{k,k+1} = \lambda_{i/n} \mbox{ , } q_{k,k-1}= \mu_{ij} \mbox{ , } qk,k = - \left( \lambda_{i/n} + \mu_{ij} \right)	
\\
\mbox{ for all }  k \geq 1
\end{split}
\]
with an invariant distribution $ \pi $, where 
\begin{equation} 
\label{eq:2}
\pi_k q_{k,k+1} = \pi_{k+1} q_{k+1,k} \mbox{ for all } k, k+1	
\end{equation}
along with the normalisation condition
\begin{equation} 
\label{eq:3}
\sum\limits_{k=0}^\infty \pi_k = 1
\end{equation} 
We obtain from (\ref{eq:2}) that 
\[
\pi_k = \left( \lambda_{i/n} / \mu_{ij} \right) \pi_{k-1}  \mbox{ for all } k \geq 1
\]
Denoting $ \left( \lambda_{i/n} / \mu_{ij} \right) $ by $ \rho_{ij} $ , we  get
\begin{equation} 
\label{eq:5}
\pi_k = \rho^k_{ij} \pi_0  	
\end{equation} 
Substituting in (\ref{eq:3}), we get 
\begin{equation} 
\label{eq:6}
\pi_k = \left( 1 - \rho_{ij} \right) \rho_{ij}^k  \mbox{ , } k = 0, 1, 2, ..., \mbox{ if } \rho_{ij} < 1
\end{equation}
which represents the invariant distribution of the Markov process transition rate of our imposed problem.

\subsubsection{Quality Model}
The markovian analytical model allows to estimate the quality of service stability. Given the expected workload $ \lambda $, the number of PMs $ m $, VMs, and the capacity of both of them, the model approximates different quality attributes; such as response time ($R$), mean queue ($W$), throughput ($T$), utilisation ($\rho$), cost ($C$) and energy consumption ($E$). 

For the VM$_{ij}$, given the incoming rate of requests $ \lambda_{i/n} $ and the mean service time $ 1 / \mu_{ij} $, the invariant queue length distribution computed in (\ref{eq:6}) gives us
\[
P(N = n) = (1 - \rho) \rho^n \mbox{, } n = 0,1,2,3, ... 
\]
In particular, $ P(N = 0) = 1 - \rho $, that is the probability that the queue is empty is steady state. Hence, the utilisation of the VM$_{ij}$ should be:  
\[
\rho_{ij} = \lambda_{i/n} / \mu_{ij}
\]
Therefore, the probability for VM$_{ij}$ to be idle can be expressed by $ \pi_0 $ from (\ref{eq:5}) as: 
\[
\pi_0 = 1 - \rho_{ij} = 1 - ( \lambda_{i/n} / \mu_{ij} )
\]

By applying Little's law $ E(S) = (1/\mu ) / (1 - \rho) $, the following performance metrics could be deduced:
\\
The mean response time for VM$_{ij}$ is estimated by:
\[
R_{ij} = 1 / (\mu_{ij} (1- \rho_{ij} )) = 1 / ( \mu_{ij} \rho_{ij} )  
\]
The mean queue length is:
\[
W_{ij} = \rho_{ij}^2 / ( 1 - \rho_{ij} )
\]
The mean throughput is basically the departure rate; i.e. the rate at which the requests finish being processed successfully at the VM; that is:
\[
T_{ij} = \lambda_{i/n} \pi_k / \sum\limits_{k=0}^\infty \lambda_{i/n} \pi_k 
\]

Having performance metrics of each VM independently, all performance metrics for a given PM could be deduced, as well as for the self-adaptive computing node. The mean response time for PM$_i$ is the mean response time for the $n_i$ VMs running on that PM. Also, the mean utilisation and the throughput can be calculated as the sum of the related measures of the $n_i$ VMs. 

On the node level, same metrics could also be calculated as the sum of related metrics for the $m$ PMs operating on the node. Operational cost could also be calculated among the node, that is the cost of processing the incoming workload: 
\[
C = \sum\limits_{i=1}^m \sum\limits_{j=1}^n Cost(CPU)_{ij}+ Cost(memory)_{ij}
\]
And, the total power consumption of all running PMs, given the varying number of VMs and their allocated CPU threads, would be:
\[
C = \sum\limits_{i=1}^m E_i
\]

As an architectural tactic represents codified knowledge about the relationship between architectural decisions and quality attributes \cite{Bachmann2002}, our analytical model can accommodate the impact of a diverse range of tactics on the stability of these quality attributes, as follows.

 \begin{itemize}
 \item Tactics related to PMs, such as horizontal scaling and consolidation, are reflected on our model by varying the value of $ m $ PMs. That is, scaling with a certain number of PMs will be reflected in our model when dividing the incoming workload $ \lambda $ on more PMs; i.e. $ m+1 $. This would influence the stability of performance (response time) and greenability (energy consumption). 
 
 \item Tactics related to VMs, such as vertical scaling and consolidation, are reflected in our model by increasing or decreasing the total value of $ n $ VMs. This influences the average latency of processing the incoming requests.
 
 \item Tactics related to computational capacity; i.e., CPU threads of a specific VM$_{ij}$; are reflected in the increase or decrease of the corresponding service rate $ \mu_{ij} $, and hence influence the throughput. Also, the utilisation of VMs, determined by our model, allows consolidating the less utilised VMs (e.g. $ x $ VMs are less than 10\% utilised) and re-checking the performance metrics given the new number of VMs ($ n-x $).
 \end{itemize}

Aiming to stabilise a certain quality attribute, the impact of related tactic could be predicted under different configurations of the tactic, in order to select the optimal configuration. Unlike prior related work, which considered a case of homogeneity, we consider the heterogeneity of environment in PMs, VMs, and their computational capacity. The proposed model is capable to model the sensitivity of quality parameters behaviour with different scenarios varying number of PMs, computational capacities of PMs, number of VMs, allocated CPU threads and requests constraints. Besides, our model allows measuring the cost and energy consumption of the self-adaptive computing node under these different scenarios. Also, information from self-awareness capabilities are employed in our model. More specifically, we rely on the goal-awareness level in informing the adaptation process to select the adaptation tactic that converges towards the adaptation goal. This influences the deduced performance metrics, and consequently leas to the choice of the optimal tactics.

\section{An Evaluation of Applicability}
\label{sec_evaluation}
We show the applicability of the proposed approach through the case of cloud architectures. First, we briefly introduce the architecture's domain, then apply the proposed work. 

Cloud-based software architectures are a suitable example of dynamism, unpredictability and uncertainty \cite{Armbrust2010}. The execution environment of cloud architectures is highly dynamic, due to the on-demand nature of the cloud. Cloud architectures operate under continuous changing conditions, e.g. changes in workload (number/size of requests), end-user quality requirements, unexpected circumstances of execution (peak demand) \cite{Chen2014a} \cite{Villegas2017}. The on-demand service provision in clouds imposes performance unpredictability and makes the elasticity of resources an operational requirement. 

Due to the on-demand and dynamic nature of cloud, there is an increasing demand on cloud services, where the realisation of quality requirements should be managed without human interventions. This type of architecture tends to highly leverage on adaptation (e.g. changing behaviour, reconfiguration, provisioning additional resources, redeployment) to regulate the satisfaction of end-users' requirements under the changing contexts of execution \cite{Salehie2009} \cite{Villegas2017}. The self-adaptation process is meant to make the system behaviour converges towards the intended behaviour, i.e. quality requirements of the end-users without SLA violation \cite{Villegas2017}. The purpose of adaptation is to satisfy the runtime demand of multi-tenant users, by changing configuration and choosing optimal tactics for adaptation. An unstable architecture will risk not improving or even degrading the system to unacceptable states \cite{Villegas2011}. In such case, there are more dynamics to observe, and stability is challenging with the continuous runtime adaptations in response to the perception of the execution environment and the system itself \cite{Villegas2017}. 

Further, the economic model of clouds (pay-as-you-go) imposes on providers economic challenges for SLA profit maximisation by reducing their operational costs \cite{Armbrust2010}. Also, providers face monetary penalties in case of SLA violations affecting their profit, which push them towards stabilising the quality of service provisioned. With the rising demand of energy, increasing use of IT systems and potentially negative effects on the environment, the environmental aspect, in terms of energy consumption, has emerged as a factor affecting the software quality and sustainability \cite{Lago2015}. While sometimes imposed by laws and regulations, decreasing energy consumption does not have only potential financial savings, but also affects the ecological environment and the human welfare \cite{Lago2015}. So, environmental requirements should be considered and traded off against business requirements and financial constraints \cite{Lago2015}. 

\subsection{Architecture Instantiation}
We instantiated the architecture of a cloud node using the reference architecture to perform stability-driven adaptations. To this end, this architecture should dynamically perform architecture-based adaptation, which would use the knowledge available at different levels of awareness in choosing optimal tactics to meet stability requirements during runtime. The instantiated architecture pattern embeds stimulus- and goal-awareness components and disables the other awareness components, to cover stability requirements of a single cloud node and focus on the evaluation of our architecture.

The stability attributes, to be taken into consideration in this case (as defined in \cite{Chen2014a}, \cite{Salama2017b}), include: (i) quality requirements specified in end-users SLAs, (ii) environmental restrictions, (iii) economic constraints, and (iv) quality of adaptation. Table \ref{tbl_StabilityAttributes} lists details of the stability attributes. With respect to the quality requirements, we consider performance (measured by response time from the time the user submits the request till the cloud submits the response back to the user in milliseconds). For the environmental aspect, we use the greenability property \cite{Lago2015} \cite{Calero2015} measured by energy consumption in kWh. For the economic constraints, we define the operational cost by the cost of computational resources (CPUs, memory, storage and bandwidth). Regarding the quality of adaptation to avoid performance degradation, we consider the settling time \textendash that is the time required by the adaptation system to achieve the adaptation goal to assure stable provision of attributes \cite{Villegas2011}. The objective functions are defined to be challenging.

\begin{table}[!h]
\caption{Stability attributes}
\label{tbl_StabilityAttributes}
\center
\footnotesize
\begin{tabularx}{0.95\textwidth}
{
>{\raggedright}p{0.17\textwidth} 
>{\raggedright}p{0.30\textwidth} 
>{\raggedright}p{0.10\textwidth} 
>{\raggedright}p{0.10\textwidth} 
>{\raggedright\arraybackslash}p{0.10\textwidth} 
}
\toprule
	\textbf{Attribute}						&
	\textbf{Description}					&	
	\textbf{Weight}							&
	\textbf{Metric}								&	
	\textbf{Objective}				
	\\
	\midrule
Performance						&
Response time						&
0.50										&
ms											&
25	
\\
Greenability							&
amount of energy consumed for operating hosts 		&
0.20										&
kWh										&
25 
\\
Operational cost					&
cost of computational resources (CPUs, memory, storage, bandwidth) 													&
0.20										&
\$												&
50		
\\
Quality of adaptation					&
time required by the adaptation system to achieve stability goals		&
0.10												&
ms													&
$ < $ 300	
\\
	\bottomrule
\end{tabularx}
\end{table}

We define the catalogue of architectural tactics to fulfil the stability attributes subject to consideration. Table \ref{tbl_QualityTactics} lists the tactics and their definitions. We base this work on the description tactics by Bass et al. \cite{Bass2012}. The tactics include: (i) horizontal scaling (increasing/decreasing the number of physical machines), (ii) vertical scaling (increasing/decreasing the number of virtual machines or their CPU capacities), (iii) virtual machines consolidation (running the virtual machines on less number of physical machines for energy savings), (iv) concurrency (by processing different streams of events on different threads or by creating additional threads to process different sets of activities), (v) dynamic priority scheduling (scheduling policy is implemented, where the scheduler handles requests according to a scheduling policy), and (vi) energy monitoring (providing detailed energy consumption information). Adaptation rules, embedded in the stimulus-awareness component, are defined as such tactics related with stability attributes. Adaptation rules are illustrated in Table \ref{tbl_AdaptationRules}.

\begin{landscape}
\begin{table}[!h]
\caption{QoS tactics and their definitions}
\label{tbl_QualityTactics}
\center
\footnotesize
\begin{tabularx}{1.40\textwidth}
{
l
>{\raggedright}p{0.20\textwidth} 
>{\raggedright}p{0.30\textwidth} 
>{\raggedright}p{0.15\textwidth} 
>{\raggedright}p{0.25\textwidth} 
>{\raggedright\arraybackslash}p{0.30\textwidth} 
}
	\toprule
	\textbf{No.}						&
	\textbf{Tactic}					&	
	\textbf{Description}		&	
	\textbf{Object}				&
	\textbf{Limits}					&	
	\textbf{Variations}			
	\\
	\midrule
1											&
Vertical scaling 				&
increasing the number of virtual machines (VMs) or their CPU capacities							&
VMs									&
maximum CPU capacity of hosts running in the datacenter		&
+1, 2, 3,... VMs or increase the CPU capacity of running VMs	
\\
2											&
Vertical de-scaling 				&
decreasing the number of virtual machines (VMs) or their CPU capacities									&
VMs											&
minimum one running VM		&
+1, 2, 3,... VMs						
\\
3											&
Horizontal scaling			&
increasing the number of running hosts					&
Hosts									&
maximum number of hosts in the datacenter			&
+1, 2, 3,... hosts				
\\
4											&
Horizontal de-scaling					&
decreasing the number of running hosts					&
Hosts													&
minimum one running host 			&
-1, 2, 3,... hosts								
\\
5											&
VMs consolidation			&
shut down hosts running least number of VMs and migrate their VMs to other hosts									&
Hosts, VMs												&
minimum one running host and one VM 			&	 
-1, 2, 3,... hosts										
\\
6															&
Concurrency 									&
processing different streams of events on different threads or by creating additional threads to process different sets of activities		&
datacenter scheduler					&
maximum CPU capacity of hosts running in the datacenter					&
single, multiple threads					
\\
7												&
Dynamic scheduling			&
scheduling policy is implemented, where the scheduler handles requests according to a scheduling policy						&
datacenter scheduler						&
maximum number of running hosts and VMs					&
earliest deadline first scheduling, least slack time scheduling, single queueing, multiple queueing, multiple dynamic queueing				
\\
\bottomrule
\end{tabularx}
\end{table}
\end{landscape}

\begin{table}[!h]
\caption{Adaptation Rules}
\label{tbl_AdaptationRules}
\center
\footnotesize
\begin{tabular}
{ lll }
	\toprule
	\textbf{Tactic}												&	
	\textbf{Related Quality Attributes}		&
	\textbf{Priority}
	\\
	\midrule
Dynamic scheduling						&
response time, throughput			&
1
\\
Conucrrency 									&
response time, throughput			&
2
\\
Vertical scaling 									&
response time, throughput				&
3
\\
Horizontal scaling							&
response time, throughput			&
4
\\
VMs consolidation												&
operational cost, energy consumption			&
1
\\
Vertical de-scaling 												&
operational cost, energy consumption				&
2
\\
Horizontal de-scaling										&
operational cost, energy consumption			&
3
\\
	\bottomrule
\end{tabular}
\end{table}

\begin{figure}[!h]
\centering
\includegraphics[width=\textwidth]{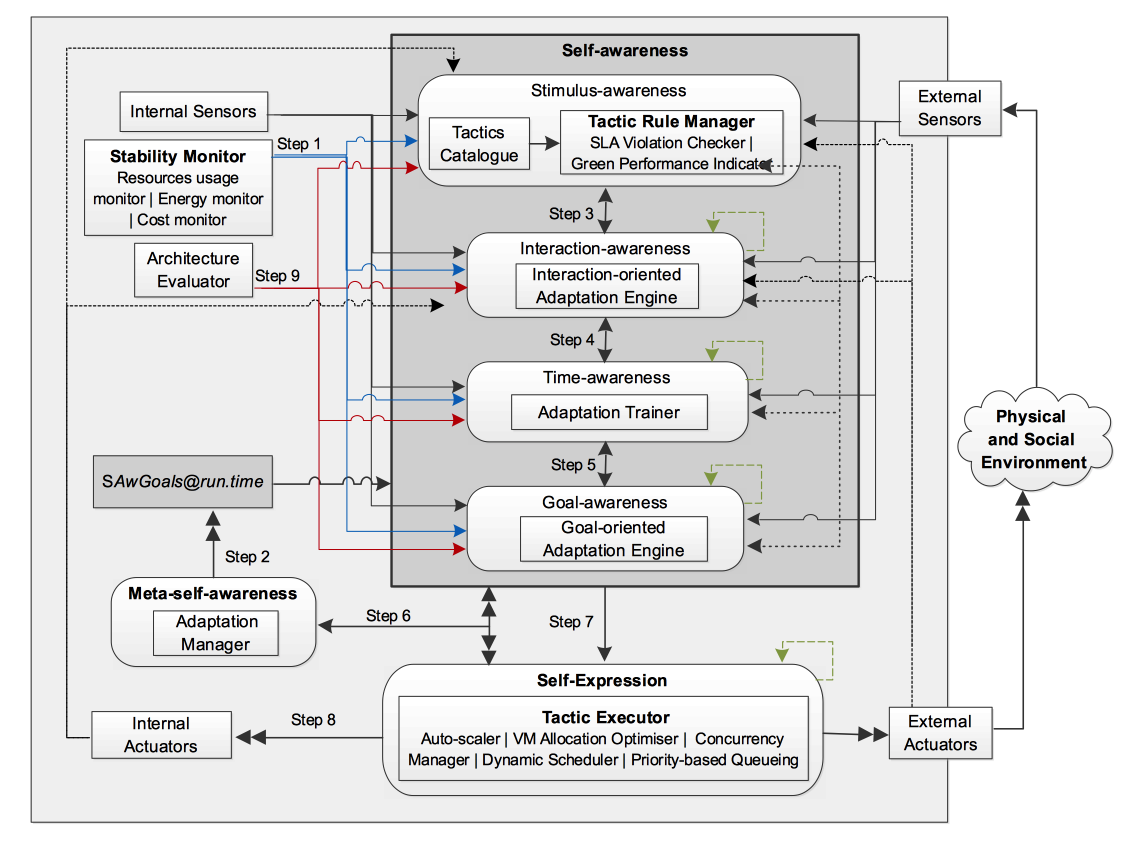}
\caption{Cloud architecture instantiated using quality-driven self-aware and self-expressive pattern}
\label{fig_cloudArchitecture}
\end{figure}

We embed the tactics catalogue and generic components in the self-awareness components and the relationships are made implicit within the interaction between different components. The architecture of the cloud node is illustrated in Figure \ref{fig_cloudArchitecture}. Tactics are defined in the Tactics Catalogue component. Monitors for stability attributes are implemented in the QoS Monitor component. Components necessary for checking possible violation of stability attributes are implemented in the stimulus-awareness component, e.g. SLA Violation Checker and Green Performance Indicator. The scheduler component of the scheduling tactic was embedded into the stimulus-aware. Management components of tactics were configured into the Tactic Executor for running the tactics, e.g. auto-scaler. 

\subsection{Modelling Stability Goals}
We define, hereunder, stability goals and runtime tactics determined above using our runtime goals modelling. Then, we provide an example of a runtime goal instance.

Stability goals {\fontfamily{lmss}\selectfont Performance} and  {\fontfamily{lmss}\selectfont QualityOfAdaptation} are dedined as follows.

\smallskip
\begin{minipage}{\linewidth}
\begin{adjustwidth}{4em}{2em}
\footnotesize
\fontfamily{lmss}\selectfont
\noindent\rule{10.5cm}{0.8pt}

\textbf{Goal} Achieve [Performance] 
	
\hangindent=2em \textbf{Informal Definition} 
		\begin{adjustwidth}{4em}{6em}
		\textit{For every request received, the request processing should be accomplished within the performance parameters defined in the SLA of the client issuing the request.} 
		\end{adjustwidth}
		
\hangindent=2em \textbf{Formal Definition} 
	
	\hspace*{2em} $ \forall $ r:Request, c:Client 
	
	\hspace*{4em} ExecuteRequest (r) $ \Rightarrow \lozenge \leq $ c.SLA(ResponseTime)
	
\hangindent=2em \textbf{Node identifier}
	$ n_1 $:Self-awareArchitectureNode

\hangindent=2em \textbf{Weight}
	$ w = 1.0 $
		
\hangindent=2em \textbf{Metric}
	ResponseTime: Request $ \rightarrow $ Time
	
	\begin{adjustwidth}{4em}{6em} 
	\textbf{def:} the duration of processing request starting from client submitting the request till submitting the response back to the client
	\end{adjustwidth}
	
\hangindent=2em \textbf{Objective functions}

	\hspace*{2em} ResponseTime $ = $ ResponseTime $ \leq $ c.SLA(ResponseTime)
	
\hangindent=2em \textbf{Tactics}
	\hspace*{0.60em} T$_1$: VerticalScaling
	
	\hspace*{4em} T$_3$: HorizontalScaling
	
	\hspace*{4em} T$_6$: Concurrency
	
	\hspace*{4em} T$_7$: DynamicScheduling

\noindent\rule{10.5cm}{0.4pt}
\newline
\end{adjustwidth}
\end{minipage}

\smallskip
\begin{minipage}{\linewidth}
\begin{adjustwidth}{4em}{2em}
\footnotesize
\fontfamily{lmss}\selectfont
\noindent\rule{10.5cm}{0.8pt}

\textbf{Goal} Achieve [QualityOfAdaptation] 
		
\hangindent=2em \textbf{Informal Definition} 
		\begin{adjustwidth}{4em}{6em}
		\textit{Any quality attribute should not be worse than 20\% of the threshold in SLA for more than 300 seconds.} 
		\end{adjustwidth}
		
\hangindent=2em \textbf{Formal Definition} 
	
	\hspace*{2em} $ \forall $ r:Request, c:Client 
	
	\hspace*{4em} QualityAttributes(r) $ \Rightarrow \lozenge _{5min} \leq 20\% $ c.SLA(QualityAttributes)
	
\hangindent=2em \textbf{Node identifier}
	$ n_1 $:Self-awareArchitectureNode
	
\hangindent=2em \textbf{Weight}
	$ w = 0.7 $
		
\hangindent=2em \textbf{Metric}
	QualityAttribute: Request $ \rightarrow $ Time
	
	\begin{adjustwidth}{4em}{6em} 
	\textbf{def:} the quality attributes of processing requests should not be worse than 20\% of the threshold in the client SLA for more than 300ms.
	\end{adjustwidth}
	
\hangindent=2em \textbf{Objective functions}

	\hspace*{2em} ResponseTime $ = $ ResponseTime $ \lozenge _{300sec} \leq 20\% $ c.SLA(ResponseTime)

\hangindent=2em \textbf{Tactics}
	\hspace*{0.60em} T$_1$: VerticalScaling
	
	\hspace*{4em} T$_3$: HorizontalScaling
	
	\hspace*{4em} T$_6$: Concurrency
	
	\hspace*{4em} T$_7$: DynamicScheduling
	
\noindent\rule{10.5cm}{0.4pt}
\newline
\end{adjustwidth}
\end{minipage}
\smallskip

Runtime tactics {\fontfamily{lmss}\selectfont VerticalScaling} and  {\fontfamily{lmss}\selectfont VMsConsolidation} are dedined as follows.

\smallskip
\begin{minipage}{\linewidth}
\begin{adjustwidth}{4em}{2em}
\footnotesize
\fontfamily{lmss}\selectfont
\noindent\rule{10.5cm}{0.8pt}

\textbf{Tactic} VerticalScaling

\hangindent=2em \textbf{Unique identifier} T1
		
\hangindent=2em \textbf{Informal Definition} 
		\begin{adjustwidth}{4em}{6em}
		\textit{increase the number of VMs or their capacities} 
		\end{adjustwidth}
		
\hangindent=2em \textbf{Object} VMs

\hangindent=2em \textbf{Pre-condition}

	\hspace*{2em} $ TotalCPUcapacity $ of running VMs $ \leq $ 
	
	\hspace*{4em} $ TotalCPUcapacity $ of hosts running in the datacenter
	
\hangindent=2em \textbf{Limits}
	$ max(TotalCPUcapacity) $ of hosts running in the datacenter
	
\hangindent=2em \textbf{Functionality}

	\hspace*{2em} increaseCPUCapacity(vm: VM) $ \vee $
	
	\hspace*{2em} increaseCoresNum(vm: VM) $ \vee $
	
	\hspace*{2em} runNewVM()
	
\hangindent=2em \textbf{Post-condition}
	Waiting Requests are migrated to the new VM
	
\hangindent=2em \textbf{Variations}
	\hspace*{0.60em} T$_{1.1}$: increase CPU capacity of 1 running VM
	
	\hspace*{5.2em} T$_{1.2}$: increase the number of cores of 1 running VM

	\hspace*{5.2em} T$_{1.3}$: add 1 VM to running VMs
	
\noindent\rule{10.5cm}{0.4pt}
\newline
\end{adjustwidth}
\end{minipage}

\smallskip
\begin{minipage}{\linewidth}
\begin{adjustwidth}{4em}{2em}
\footnotesize
\fontfamily{lmss}\selectfont
\noindent\rule{10.5cm}{0.8pt}

\textbf{Tactic} VMsConsolidation

\hangindent=2em \textbf{Unique identifier} T4
		
\hangindent=2em \textbf{Informal Definition} 
		\begin{adjustwidth}{4em}{6em}
		\textit{shut down hosts running least number of VMs and migrate their VMs to other hosts} 
		\end{adjustwidth}
		
\hangindent=2em \textbf{Object} Hosts, VMs

\hangindent=2em \textbf{Pre-condition}
	$ number $ of hosts running in the datacenter $ \geq 2 $ 
	
\hangindent=2em \textbf{Limits}
	\hspace*{0.60em}  $ min $ 1 host running in the datacenter $ \wedge $
	
	\hspace*{3.6em} $ min $ 1 VM running 
	
\hangindent=2em \textbf{Functionality}

	\hspace*{2em} migrateVMs(host: Host) $ \vee $
	
	\hspace*{2em} shutdown(host: Host)
		
\hangindent=2em \textbf{Post-condition}
	Requests are migrated to VMs
	
\hangindent=2em \textbf{Variations}
	\hspace*{0.60em} T$_{4.1}$: shutdown 1 host	
	
\noindent\rule{10.5cm}{0.4pt}
\newline
\end{adjustwidth}
\end{minipage}
\smallskip

An instance of the Runtime Goal {\fontfamily{lmss}\selectfont Performance} is defined as follows. 

\bigskip
\begin{minipage}{\linewidth}
\begin{adjustwidth}{4em}{6em}
\footnotesize
\fontfamily{lmss}\selectfont
\noindent\rule{9.5cm}{0.8pt}

\textbf{Goal} $ G_1(n_1, t_i) $
		
\hangindent=2em \textbf{Client} c:Client 

\hangindent=2em \textbf{Request} r:Request 

\hangindent=2em \textbf{ObjectiveFunction} ResponseTime$_c = $ ResponseTime(r) $ \leq 15ms $ 

\hangindent=2em \textbf{AdaptationAction} T$_{1.3}$

\hangindent=2em \textbf{SatisfactionDegree} $ v = 14ms $ 

\hangindent=2em \textbf{Environment Runtime Goals} $ G_e(t_i) = \{G_1(N_2, t_i), G_1(N_3, t_i), ...\} $ 

\hangindent=2em \textbf{Environment Runtime Tactics} $ T_e(t_i) = \{N_2.T_{4.1}, N_3.T_{1.3}\} $ 

\noindent\rule{9.5cm}{0.4pt}
\newline
\end{adjustwidth}
\end{minipage}

\subsection{Experimental Evaluation}
\label{sec_evaluation_}
The main objective of the experimental evaluation is to examine the stability attributes when using the instantiated architecture and goals modelling, and assess associated overhead. 

\subsubsection{Experiments Setup}
\label{sec_evaluation_setup}
To conduct the experimental evaluation, we implemented the instantiated architecture using the widely adopted \textit{CloudSim} simulation platform for cloud environments \cite{Calheiros2011}. The simulation was built using Java JDK 1.8, and was run on a 2.9 GHz Intel Core i5 16 GB RAM computer. We set the runtime goals model with with stability attributes as defined in Table \ref{tbl_StabilityAttributes}. We configured adaptation tactics as defined in Table \ref{tbl_QualityTactics}, and configured different self-awareness components to use the adaptation rules defined in Table \ref{tbl_AdaptationRules}.

We used benchmarks to stress the architecture with highly frequent changing demand and observe stability goals. To simulate runtime dynamics, we used the \textit{RUBiS} benchmark \cite{RUBiS} and the \textit{World Cup 1998} trend \cite{WorldCup98} in our experiments. The \textit{RUBiS} benchmark \cite{RUBiS} is an online auction application defining different services categorised in two workload patterns: the browsing pattern (read-only services, e.g. BrowseCategories), and the bidding pattern (read and write intensive services, e.g. PutBid, RegisterItem, RegisterUser). For fitting the simulation parameters, we mapped the different services of the \textit{RUBiS} benchmark into Million Instructions Per Second (MIPS), as listed in Table \ref{tbl_experimentsServices}. To simulate a realistic workload within the capacity of our testbed, we varied the number of requests proportionally according to the \textit{World Cup 1998} workload trend \cite{WorldCup98}. We compressed the trend in a way that the fluctuation of one day (=86400 sec) in the trend corresponds to one time instance of 864 seconds in our experiments. This setup can generate up to 700 parallel requests during one time instance, which is large enough to challenge stability. 

\begin{table}[!h]
\caption{Types of service requests}
\label{tbl_experimentsServices}
\center
\footnotesize
\begin{tabularx}{0.75\textwidth}
{
		>{\raggedright}p{0.27\textwidth} 
		l
		l
		l
}
\toprule
{\textbf{Service Pattern}} 		& 
{\textbf{S\#}} 								& 
{\textbf{Service Type}} 			& 
{\textbf{Required MIPS}}
\\
\midrule
browsing only			&
1									&
read-only					&
10,000  
\\
\hline
bidding only				&
2									&
read and write			&
20,000  
\\
\hline
\multirow{3}{*}{\parbox{3cm}{mixed with adjustable composition of the two service patterns}}															&
3																											&
70\% browsing, 30\% bidding												&
12,000  
\\
																											&
4																											&
50\% browsing, 50\% bidding												&
15,000  
\\
																											&
5																											&
30\% browsing, 70\% bidding												&
17,000  
\\
\bottomrule
\end{tabularx}
\end{table}

The initial deployment of the experiments is 10 running hosts IBM x3550 server, each with the configuration of 2 x Xeon X5675 3067 MHz, 6 cores and 256 GB RAM. The frequency of the servers' CPUs is mapped onto MIPS ratings: 3067 MIPS each core \cite{Beloglazov2012} and their energy consumption is calculated using power models of \cite{Beloglazov2012}. The maximum capacity of the architecture is 1000 hosts. The characteristics of the virtual machines (VMs) types correspond to the latest generation of General Purpose Amazon EC2 Instances \cite{AmazonEC2}. In particular, we use the m4.large (2 core vCPU 2.4 GHz, 8 GB RAM), m4.xlarge (4 core vCPU 2.4 GHz, 16 GB RAM), and m4.2xlarge (8 core vCPU 2.4 GHz, 32 GB RAM) instances. The operational cost of different VMs types is 0.1, 0.2 and 0.4 \$/hour respectively. Initially, the VMs are allocated according to the resource requirements of the VM types. However, VMs utilise less resources according to the workload data during runtime, creating opportunities for dynamic consolidation. The initial deployment of the experiment is shown in Table \ref{tbl_experimentsConfiguration}.

\begin{table}[!h]
\caption{Initial deployments of the experiments}
\label{tbl_experimentsConfiguration}
\center
\footnotesize
\begin{tabularx}{0.55\textwidth}
{
	l
	>{\raggedright\arraybackslash}p{0.75\textwidth} 
}
\toprule
{\textbf{Configuration}} 		& 	
{\textbf{}}					
\\
\midrule
Hosts type		&
IBM x3550 server
\\
Hosts Specs	&
2 x Xeon X5675 3067 MHz, \newline 6 cxores, 256 GB RAM
\\
\hline
VMs types		&
General Purpose Amazon EC2 Instances 
\\
VMs Specs			&
m4.large: 2 core CPU 8 GB RAM	\newline
m4.xlarge: 4 core CPU 16 GB RAM 	\newline
m4.2xlarge: 8 core CPU 32 GB RAM
\\
\hline
No. of hosts			&
10 (max. 1000)		
\\
No. of VMs						&	
5 x m4.large,	 5 x m4.xlarge,	 5 x m4.2xlarge
\\
\bottomrule
\end{tabularx}
\end{table}

\subsubsection{Results of Stability Attributes}
\label{sec_evaluation_ResultsStabilityAttributes}
We report, first, on the average of stability attributes at each time interval. We examined stability attributes at each time interval of 864 seconds. More specifically, we run the entire workload for each service type and measured the stability attributes in our self-aware architecture compared to self-adaptive architecture. The implemented self-adaptive architecture is a self-adaptive MAPE architecture \cite{Maurer2011}.

\begin{figure*}[!h]
\centering
\includegraphics[width=\textwidth]{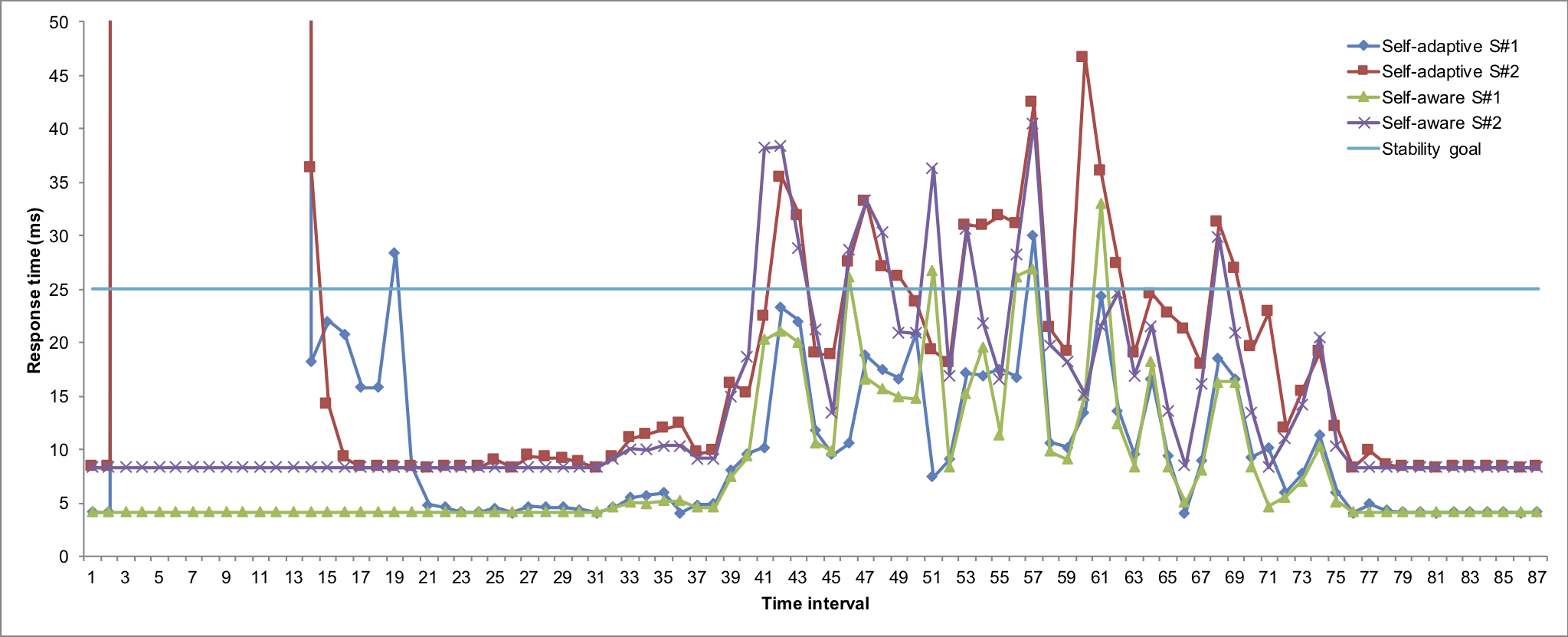}
\caption{Average response time in time intervals}
\label{graph_resultsResponseTime}
\end{figure*}

\begin{figure*}[!h]
\centering
\includegraphics[width=\textwidth]{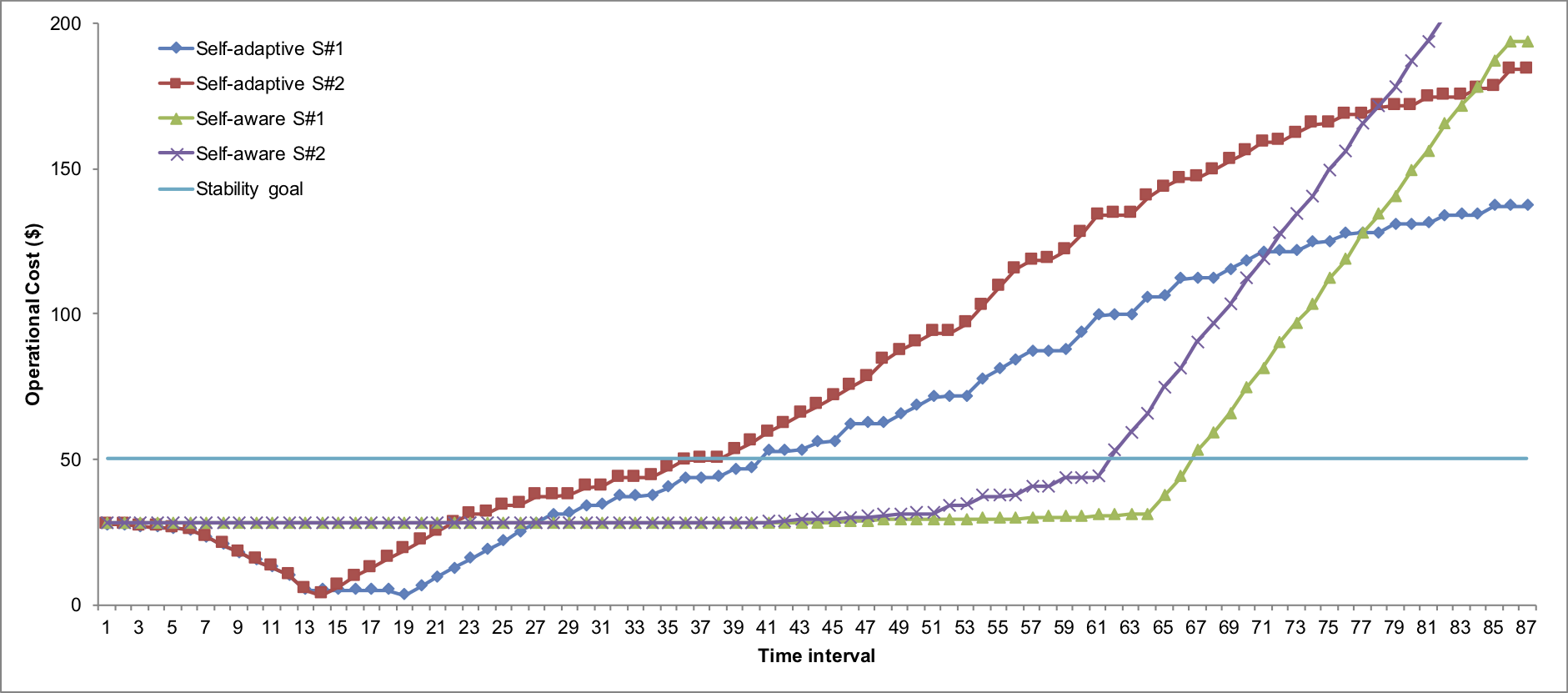}
\caption{Average operational cost in time intervals}
\label{graph_resultsCost}
\end{figure*}

The results of response time and operational costs for service types 1 and 2 are depicted in Figure \ref{graph_resultsResponseTime} and \ref{graph_resultsCost}. As shown in the figures, the self-aware architecture was able to result more stable response time compared with the self-adaptive architecture in both service types, while the latter caused violation in response time in early time intervals when the peak workload started. At the same time, the self-aware architecture was also capable to stabilise the operational cost for longer time intervals than the self-adaptive architecture. It is worth noting that stabilising both response time and cost at the same time is very challenging in case of peak workload, that is why the self-aware architecture considered keeping the response time without violations while not fully stabilising the cost within the constraint while keeping the response time without violations, as per the response time weight is higher. Yet, achieving stability of both for longer time intervals reflects the higher quality of adaptations and tactics selection. 

Taking a closer look at the service requests, Table \ref{tbl_resultsViolations} shows the violation in response time for all service types and associated quality of adaptation. The quality of adaptation is calculated, here, by the time periods where the response time was violated. As shown in the table, the percentage of violations in response time is slightly better in the case of self-aware architecture in all service types. But regarding the total periods of time where the response time was worse than 20\% of the SLA constraint, the self-aware architecture was capable to keep it much less than the self-adaptive architecture. For instance, response time violation in the case of service type 1 was 12.96\% and 14.23\% for the self-adaptive and self-aware architectures respectively, while the total time of violations was 11232 sec compared to 4320 sec. Meanwhile, the self-aware architecture violations were less for service types 2 and 4, with better quality of adaptation.

\begin{table}[!h]
\caption{Average stability goals of all service requests}
\label{tbl_resultsViolations}
\center
\footnotesize
\begin{tabularx}{0.60\textwidth}
		{	l l l l l  }
\toprule
{\textbf{S\#}}							&
\multicolumn{2}{l}{\textbf{Response time violation (\%)}} 		&
\multicolumn{2}{l}{\textbf{Quality of adaptation (sec)}} 
\\
										&
Self-adaptive				&	
Self-aware					&
Self-adaptive				&	
Self-aware
\\
\midrule
1 									&	
12.96							&	
14.23							&
11232							&
4320
\\
2									&	
26.44							&	
24.40							&
24192							&
9504	
\\
3									&	
15.78							&	
17.19							&
13824							&
5184							
\\
4									&	
20.90							&	
20.91							&
19872							&
6048
\\
5									&	
27.51							&	
28.31							&
19008						&
8640
\\
\bottomrule
\end{tabularx}
\end{table}

Considering the experiments total results, we report the average of 30 runs in Table \ref{tbl_experimentsResults}. Adaptation overhead is calculated by the time spent in the adaptation process. The average response time of all requests for each service type is much better achieved by the self-aware architecture (average 62.85 ms compared to 20.02 ms). This came with the price of higher operational cost (168.94 \$ vs. 205.84 \$) and adaptation overhead (average 164.90 sec vs. 251.62 sec). Yet, the difference in response time is much bigger than the difference in cost and overhead. While the energy consumption in self-adaptive architecture was less (24.96 kWh), the self-aware architecture was capable to keep it within the stability goal (28.52 khW).

\begin{table}[!h]
\caption{Experiments total results}
\label{tbl_experimentsResults}
\center
\footnotesize
\begin{tabularx}{0.70\textwidth}
		{	
		>{\raggedright}p{0.35\textwidth} 
		l
		l 
		l 
		}
\toprule								
\textbf{Experiments results	}		&
{\textbf{S\#}}									&
\multicolumn{2}{c}{\textbf{Architecture Pattern}} 
\\
										&
										&
Self-adaptive				&	
Self-aware
\\
\midrule
Average response time (ms)				&	
1															&	
73.73													&	
16.00
\\
										&	
2										&	
63.49								&	
22.92
\\
			 							&	
3										&	
58.41								&	
18.56
\\
										&	
4										&
58.90								&	
21.10
\\
										&	
5										&	
59.74								&	
21.54
\\
										&
avg.								&
62.85								&
20.02
\\
\hline
Average energy consumption (kWh)				&	
1																&	
23.80														&	
28.52
\\
			 							&	
2										&	
25.33								&	
28.52
\\
									&	
3									&
25.02							&	
28.52
\\
									&	
4									&	
25.33							&	
28.52
\\
									&	
5									&	
25.33							&	
28.52
\\
										&
avg.								&
24.96								&
28.52
\\
\hline
Total operational cost (\$)			 							&	
1														&	
137.18											&	
193.44
\\
								&	
2								&
184.08					&	
230.88
\\
								&	
3								&	
159.02					&	
199.38
\\
								&	
4								&	
180.66					&	
199.68
\\
										&
avg.								&
168.94							&
205.84
\\
\hline
Adaptation overhead (sec)		 		&	
1																&	
157.80													&	
247.40
\\
								&	
2								&
168.50					&	
260.60
\\
								&	
3								&	
162.70					&	
249.00
\\
								&	
4								&	
167.40					&	
249.60
\\
			 					&	
5								&	
168.10					&	
251.50
\\
										&
avg.								&
164.90							&
251.62
\\
\bottomrule
\end{tabularx}
\end{table}

\subsection{Discussion}
\label{sec_evaluation_discussion}
The proposed architecture having generic components to embed runtime tactics have successfully instantiated many tactics for different quality and stability attributes and enriched the self-aware patterns with self-management quality capabilities to meet the changing workload and stabilise quality requirements during runtime. Evaluating the features of the proposed approach and reference architecture is summarised as follows:

\begin{itemize}
\item \textit{Efficiency.} The ability to incorporate a range of tactics for different stability attributes into the patterns diversify the catalogue space from which the adaptation actions could be selected and implemented during runtime to meet stability requirements under dynamic workload. 

\item \textit{Ease of instantiation and use.} The structure of the tactics for different quality attributes was embedded efficiently within the generic components of the reference architecture. Their interaction specification also took place within the process flow while taking advantage of the self-awareness knowledge available from different self-awareness levels.

\item \textit{Multiple uses.} The generic approach for instantiating the architecture allowed featuring different combinations of self-awareness capabilities. Thus, incorporating tactics approach could be used in any of these patterns according to the requirements of the system, without unnecessary overhead caused by self-awareness components.
\end{itemize}

Generally, the proposed architecture and goals modelling for stability have proven feasibility when embedding tactics for different stability attributes. The proposed architecture tends to diversify the possible adaptation actions to be taken during runtime. The quantitative evaluation has proven the ability of the architecture and goals model to efficiently realise stability and enhance the quality of adaptation.

\section{Threats to Validity}
\label{sec_threats}
The potential threats to validity of the proposed method are noted below:

\begin{itemize}
\item The dependency on the human capabilities in selecting the architecture pattern would form a threat to validity on the end results. This might be due to the lack of information or expertise knowledge. Yet, our approach could be complemented with symbiotic simulations for testing the architecture design \cite{Turner2011} \cite{Tjahjono2015}.

\item The fact that the proposed method is evaluated by its authors presents a threat to objectivity. To mitigate this risk, we sought to conduct practical evaluation by architects in industrial settings, in order to provide more feedback from independent sources.

\item Another threat to validity of our evaluation lies in the fact that the approach was evaluated using one case. Yet, the dynamics presented in cloud architectures is an appropriate case study representing dynamics of modern software systems, and we plan to conduct other case studies in industrial contexts and different business segments. 

\item Subjectivity might be considered a threat to validity in setting the stability attributes, as it was conducted based on the authors' background and knowledge. Our strategy mitigation for this issue has been basing the case study on previous work of \cite{Chen2014a} \cite{Salama2015} \cite{Salama2016} \cite{Salama2017b}, this makes us believe that the case study is practical and reflects the nature of cloud-based software systems.

\item Experiments were conducted in a controlled environment and have not considered the real-life scenario of switching between different service patterns and changing stability goals during runtime for different end-users. Given the use of a real-world workload trend and the \textit{RUBiS} benchmark, we consider that our experiments have given good enough indication and approximation of likely scenarios in a practical setting. Also, we have chosen the stability goals thresholds purely based on our observations, e.g. response time not exceeding 25 ms. Yet, these goals have proved to be challenging when running the experiments. 

\end{itemize}

\section{Related Work}
\label{sec_relatedWork}
In this section, we discuss related work in the context of architecture patterns and goals modelling.

\subsection{Architecture Patterns and Tactics}
\label{sec_relatedWork_patterns}
A large body of research in architecture design has yielded the development of approaches for incorporating and using tactics in the context of software architectures. For instance, a systematic approach for building software architecture that embodies quality requirements using architectural tactics has been proposed \cite{Kim2008} \cite{Kim2009}. Other efforts focused on tactics for certain quality attributes, such as modifiability tactics \cite{Bachmann2007}, performance tactics \cite{Champagne2011} \cite{Koziolek2011}. Others tackled the application of tactics, such as analysing the application of tactics \cite{Sanchez2011} and recommendation \cite{Mirakhorli2013}. But stability has not been explicitly considered as a property in designing software architectures.

The self-adaptive architecture community has developed in the area of quality management. For instance, the Rainbow framework \cite{Garlan2004} was proposed to support such adaptation, where strategies in the adaptation engine are architectural tactics. A framework for evaluating quality-driven self-adaptive software systems was proposed using a set of metrics to evaluate quality attributes and adaptation properties \cite{Villegas2011}. While literature has widely covered the incorporation of tactics in the context of software architectures, yet till recently, architecture patterns and tactics for self-adaptive and self-aware software have received little attention, as to the best of our knowledge \cite{Weyns2013} \cite{Chen2014a} \cite{Salama2015}. A reference architecture for self-adaptive software has been proposed based on reflection \cite{Affonso2013} but designing for stability with self-awareness has not been tackled yet.

\subsection{Goals Modelling}
\label{subsec-goals}
Related work, geared towards runtime requirements modelling, are ``models@run.time'' and ``self-explanation''. 

Models@run.time rethinks adaptation mechanisms in a self-adaptive system by leveraging on model-driven engineering approaches to the applicability at runtime \cite{Blair2009}. This approach supports requirements monitoring and control, by dynamically observing the runtime behaviour of the system during execution. Models@run.time can interleave and support runtime requirements, where requirements and goals can be observed during execution by maintaining a model of the requirements in conjunction to its realisation space. The aim is to monitor requirements satisfaction and provide support for unanticipated runtime changes by tailoring the design and/or invoking adaptation decisions which best satisfy the requirements. Meanwhile, authors in \cite{Lamsweerde2003} proposed a goal-oriented approach for systematically building architecture design from system goals.

In the context of self-adaptive systems, self-explanation was introduced to adaptive systems to offer interpretation of how a system is meeting its requirements, using goal-based requirements models \cite{Welsh2014}. Self-explanation focused mainly on explaining the self-adaptive behaviour of the running system, in terms of satisfaction of its requirements, so that developers can understand the observed adaptation behaviour and garner confidence to its stakeholders. Authors in \cite{Cavalcante2015} have theoretically revisited goal-oriented models for self-aware systems-of-systems. Goal models were also introduced as runtime entities in adaptive systems \cite{Goldsby2008} and context-aware systems \cite{Vrbaski2012}.

Though, there has been growing research in runtime requirements engineering in the context of self-adaptive software systems, yet these models and approaches have limitations in enabling the newly emerged self-properties, i.e. self-awareness and self-expression. To the best of our knowledge, there is no research that tackled goals modelling for self-aware and self-expressive software systems, as well realising the symbiotic relation between both.

\section{Conclusion and Future Work}
\label{sec_conclusion}
In this paper, we presented a reference architecture for architectural stability, using a generic approach for incorporating architecture tactics and QoS self-management components in self-aware architecture patterns. The approach is based on providing the self-aware patterns with a catalogue of architectural tactics designated to fulfil different stability attributes. The stability-based adaptation will be performed during runtime by the awareness capabilities available in different patterns. Using the case of cloud architecture, quantitative experiments have proven enhancements in achieving stability and quality of adaptation using the reference architecture and goals modelling for stability. Our future work will focus on explicit management of trade-offs between stability attributes to achieve better adaptation. We also plan to validate the proposed method in practice by implementing its elements for cloud infrastructure-as-a-service (IaaS) management software systems, such as OpenStack \cite{OpenStack}.

\begin{acks}
Thanks are due to Rodrigo N. Calheiros for his support in the simulation development using CloudSim.
\end{acks}

\bibliographystyle{ACM-Reference-Format}
\bibliography{JStabilityArchDesign-bib}

\end{document}